\documentclass[3p]{elsarticle}

\makeatletter
\def\ps@pprintTitle{%
 \let\@oddhead\@empty
 \let\@evenhead\@empty
 \def\@oddfoot{}%
 \let\@evenfoot\@oddfoot}
\makeatother

\newcommand{\red}[1]{{\color{black}#1}}

\newcommand{\mdisc}[1]{\mathbf{#1}}
\newcommand{\sdisc}[1]{#1}
\usepackage{xcolor,textcomp}
\usepackage{graphicx,hyperref,epstopdf,mathtools}
\usepackage{amsmath,amssymb,multirow,array,gensymb,bm}
\usepackage{fancyhdr,float}
\usepackage[none]{hyphenat}
\usepackage{enumitem,relsize}
\usepackage{footmisc,mathtools}
\usepackage{soul,subfigure,longtable}
\usepackage[utf8]{inputenc}
\usepackage{lineno}

\newcolumntype{P}[1]{>{\centering\arraybackslash}p{#1}}
\newcolumntype{M}[1]{>{\centering\arraybackslash}m{#1}}

\biboptions{sort&compress}

\AtBeginDocument{%
  \def\thefnote{\myfnsymbol{fnote}}}
\makeatletter
\def\myfnsymbol#1{\expandafter\@myfnsymbol\csname c@#1\endcsname}
\def\@myfnsymbol#1{\ifcase #1\or $\dagger$\or $\#$\else \@ctrerr\fi}
\def\fntext[#1]#2{\g@addto@macro\@fnotes{%
   \refstepcounter{fnote}\elsLabel{#1}%
   \def\thefootnote{\thefnote}
   \global\setcounter{footnote}{\c@fnote}%
   \footnotetext{#2}}}
\makeatother

\begin{document}

\begin{frontmatter}

\title{A Framework to Systematically Study the Nonlinear Fluid-Structure Interaction of Phononic Materials with Aerodynamic Flows}

\author[inst1]{Vinod Ramakrishnan}
\author[inst2]{Arturo Machado Burgos}
\affiliation[inst1]{organization={The Grainger College of Engineering, Department of Mechanical Science and Engineering},
  addressline={University of Illinois Urbana-Champaign},
  postcode={Illinois, 61801},
  city={Urbana},
  country={USA}}
  \affiliation[inst2]{organization={The Grainger College of Engineering, Department of Aerospace Engineering},
  addressline={University of Illinois Urbana-Champaign},
  postcode={Illinois, 61801},
  city={Urbana},
  country={USA}}
\author[inst1]{Sangwon Park}
\author[inst1]{Kathryn H. Matlack}
\author[inst2]{Andres Goza\corref{cor1}}
\cortext[cor1]{Corresponding author}
\ead{agoza@illinois.edu}

\begin{abstract}

Phononic materials (PMs) are periodic media that exhibit novel elastodynamic responses. While PMs have made progress in vibration-mitigation applications, recent studies have demonstrated the potential of PMs to passively and adaptively modulate flow behavior through fluid-structure interaction (FSI). For example, PMs have been shown to delay laminar-to-turbulent transition and mitigate unsteadiness in shock-boundary layer interactions. However, a systematic framework to relate the effect of specific PM behaviors to the FSI dynamics is lacking. Such a framework is essential to systematically investigate the complex and nonlinear coupled dynamics of the FSI. Further, parameters that are not typically considered in PM models become critical, such as the vibration amplitude. This article addresses this gap by proposing FSI-relevant ``behavioral'' parameters, distinct from the structural parameters of the PM, but with a clear mapping provided to them. We use high-fidelity, strongly coupled simulations to quantify the FSI between a novel configuration of laminar flow past a flat plate, equipped with a PM. Our study proposes four critical PM behavioral parameters---effective stiffness, truncation resonance frequency, a quantity representing the dynamic displacement amplitude, and unit cell mass---that influence the spectral characteristics of the vortex-shedding process inherent to the flat plate system. Results show connections between each parameter and distinct behavior in the lift coefficient in FSI. While the focus of this work is on the PM-FSI dynamics in an aerodynamic flow, we argue that identifying these behavioral parameters is key to unlocking scientific study and design with phononic materials in fluid flows more broadly.

\end{abstract}


\end{frontmatter}

\section{Introduction}\label{sec:Introduction}

Phononic materials (PMs) are periodic, architected media that leverage intricate micro- and/or meso-scale structural features to customize the elastodynamic macro-scale response~\cite{KochmannAMR2017,PatilAM2022,JiaoNC2023,WuAFM2024}. PM architectures can exhibit versatile, atypical behaviors such as band gaps~\cite{KushwahaPRL1993,KushwahaIJMPB1996}, directional wave-propagation~\cite{LeeSMS2023}, topological behavior~\cite{VishwakarmaCP2025}, auxetic behavior~\cite{ZhangAFM2025}, and non-reciprocity~\cite{WuJSV2019}, promoting their use in energy harvesting~\cite{AkbariSAA2023}, cloaking~\cite{MartinezMTS2022}, and morphing structures~\cite{AksoySA2022}. Recently, dynamic features of PMs, such as truncation resonances \cite{BastawrousJASA2022,HasanPRSA2019,HasanJEL2024} and defect modes~\cite{RamakrishnanJSV2025}, have been leveraged to beneficially interact with fluid flows in various flow settings~\cite{hussein2015flow,WilleyJFS23,MichelisPoF2023,BarnesAIAA2021,SchmidtJAP2025,KeoghArXiV2025,NavarroMatter2025}. Existing studies have demonstrated the effectiveness of PMs in altering Tollmien–Schlichting wave boundary layer instabilities~\cite{hussein2015flow,WilleyJFS23,MichelisPoF2023,BarnesAIAA2021,SchmidtJAP2025,KeoghArXiV2025} and shockwave/boundary-layer interactions~\cite{NavarroMatter2025}, leading to effects such as a change in the spatial location of laminar-to-turbulent transition and stabilization of hypersonic flow, respectively. Although these preliminary studies have shown promise for PMs to alter flow dynamics, a framework for systematically investigating and, ultimately, designing FSI with PMs remains elusive. This article progresses toward addressing this gap by proposing and justifying behavioral PM parameters that enable systematic tuning of the FSI dynamics.

To motivate the challenges in building this framework for PM-FSI, we refer to insights from the FSI community. Canonical FSI systems that have received particularly extensive attention include vortex-induced vibration of flow past an elastically mounted cylinder, flow past a compliant flat plate/beam/elastica in either an unbounded or channel configuration, and transitional/turbulent wall-bounded flow past a compliant wall. (There are many others involving FSI associated with structures undergoing prescribed kinematics, which we neglect here for conciseness and because our flow system of interest involves a nominally fixed aerodynamic body). We refer the reader to references \cite{riley1988compliant, Gad-el-Hak1996compliant, williamson2004vortex, sarpkaya2004critical, shelley2011flapping} for extensive reviews of a variety of these systems. Our focus here is on summarizing key outcomes of coupling flow and structure systems, and their implications for systematically studying FSI involving PMs. 

Flow-structure coupling can produce complex and possibly counterintuitive dynamics. Across a variety of FSI systems, the dynamics may be systematically varied by tuning a representative structural mass (relative to a characteristic flow mass) and structural stiffness (relative to a notion of stiffness in the flow, typically induced by the dynamic pressure) \cite{riley1988compliant, Gad-el-Hak1996compliant, williamson2004vortex, sarpkaya2004critical, shelley2011flapping}. (A common variant to this parameter set, particularly for vortex-induced vibration, is to replace the stiffness with a flow velocity relative to the structural natural frequency \cite{sarpkaya2004critical, williamson2004vortex}). Varying these mass and stiffness (velocity) parameters triggers regime changes, including the onset of instability, as well as self-induced and sustained large-amplitude vibrations characterized by either limit cycle, quasi-periodic, or chaotic behavior, to name a few possibilities \cite{riley1988compliant, Gad-el-Hak1996compliant, williamson2004vortex, sarpkaya2004critical, shelley2011flapping}. 

The onset of FSI vibrations/oscillations is often studied through a linear analysis, where the flow-structure system is linearized about a steady base state.
Even when the coupling is linearized, coupled instability modes (not present if the compliant structure is replaced with a perfectly rigid one) can arise \cite{benjamin1960effects, landahl1962stability}. Accordingly, structural designs must be assessed not by their interaction with a single mode of the linearized flow system for a rigid body; instead, the stability of the entire (linearized) FSI system after the compliant structure is incorporated must be systematically analyzed. Linear stability analysis and, more recently, resolvent analysis have been utilized for a variety of coupled FSI systems; a few representative references are \cite{benjamin1960effects, landahl1962stability, carpenter1985hydrodynamic, shelley2005heavy, doare2011piezoelectric, luhar2015framework, goza2018global-4ac, pfister2022global}. Such techniques must be adopted for the assessment and design of PMs in modulating fluid flows. Approaches along these lines have been pursued for PMs in \citet{BarnesAIAA2021, WilleyJFS23}, where the linear effect of structural motion on the flow is formally accounted for via an impedance function. \citet{SchmidtJAP2025} also demonstrated that the structural PM model could be coupled to the impedance function, to directly connect the structural motion to the PM design. 

Other dynamics are possible when nonlinear effects are fully accounted for by tuning the mass and stiffness parameters. A common behavior for the FSI system is to `lock onto' a characteristic system frequency. Traditionally, this lock-on was defined with respect to a vacuum-structural natural frequency (see, for example, reference \cite{fredsoe1997hydrodynamics}). However, studies demonstrated a capacity for an FSI system to vibrate at a single frequency far from the vacuum structural frequency, or even when the structure was perfectly massless \cite{shiels2001flow}. These outcomes led to broader definitions, based on whether flow-structure dynamics synchronized, possible due to a modified structural mass that incorporates added mass effects \cite{sarpkaya2004critical}. Other representative FSI behaviors include the capacity for chaotic vibrations, even in a laminar flow setting. For example, \citet{connell2007flapping} showed that in the flow past a flapping flag, the mass is a bifurcation parameter; with increasing value at a fixed stiffness, the system can transition from no flapping to limit-cycle flapping to chaotic flapping, which the authors argued as a period-doubling effect by the appearance of additional flapping frequencies. This non-exhaustive overview illustrates the capacity for self-induced instability, regime changes including transitions from stability to limit cycle vibrations to chaotic vibrations, the selection of dominant FSI timescales by representative system frequencies, and the importance of the governing mass and stiffness (velocity) parameters in driving this behavior. These behaviors, and more broadly insights from the FSI community, are particularly important in the context of certain PM efforts, such as designing PMs with out-of-phase behavior to the flow signatures, to reduce flow instabilities. It is important, therefore, to account for possible flow phase-alignment with structural dynamics or shifts in the frequency of the coupled system altogether. Along these lines, continually expanding analysis of PM-FSI systems to account for the nonlinear coupling, either via experiments or high-fidelity nonlinear simulations, is important. Studies such as in references \cite{BarnesAIAA2021,WilleyJFS23} have provided a platform for this approach, connecting PMs designed through a linear analysis to nonlinear simulations that resolve the nonlinear FSI coupling.

A corresponding set of PM-FSI parameters, analogous to a representative mass and stiffness for traditional FSI systems, does not exist. Indeed, a direct use of a characteristic mass and stiffness, nearly ubiquitous across typical FSI systems, is unlikely to capture the effect of PMs on FSI. For example, most PM systems are defined by a set of distinct (periodically repeating) mass and/or stiffness elements. Moreover, the key opportunity in incorporating PMs into fluid flows is the potential to leverage engineered structural behaviors, not possible with traditional homogeneous structures, to favorably alter a flow. As such, a collection of parameters that enables a systematic tuning of PM-centric behaviors and their effect on the flow is a missing central enabler to PM-FSI study and design. 

In this vein, this article proposes FSI-relevant ``behavioral'' parameters of the PM that are distinct from its structural parameters, e.g., mass and stiffness of the PM, but with a clear one-to-one mapping provided between these parameters. We test a hypothesis that these behavioral parameters will allow a systematic parameterization of the complex FSI. The effect of these PM behavioral parameters on FSI and flow characteristics is comprehensively assessed using high-fidelity, strongly coupled FSI simulation via an immersed boundary method. The computational methodology accurately treats the viscous, nonlinear separated flow processes, employs canonical reduced-order models that convey key PM behaviors of focus, and fully accounts for the nonlinear flow-structure coupling between these systems. Simulations are performed for a configuration of flow past an angled flat plate equipped with a phononic subsurface, with the plate angle of attack set to just below that where vortex shedding arises. A suite of fully coupled simulations, systematically sweeping over the behavioral parameters, is presented. For each set of parameters, we quantify the effects of the PM on the coefficient of lift, vortex-shedding features, forcing at the PM-flow interface, and the PM response from the fully coupled FSI simulations. 

Through this analysis, we argue how the behavioral parameters provide a clear pathway to systematically tune the coupled PM-aerodynamic flow system. Further, the results reveal the benefit of considering PMs for FSI over simpler, more traditional structures with fewer degrees of freedom (DOFs). Ultimately, this study aims to demonstrate the broader relevance of utilizing carefully chosen PM behavioral parameters in enabling efficient PM design for FSI settings with different flow configurations. Further, while the primary focus is on the ability to tune FSI through appropriately chosen parameters, the FSI results demonstrate the potential relevance of PMs to favorably manipulate flow over lifting bodies, for aerodynamic benefit. 

The remainder of the article is organized as follows. Sec.~\ref{sec:probmethod} briefly summarizes the FSI problem configuration. Sec.~\ref{sec:PM} provides an overview of the dynamic analysis of PM models considered for this study, and proposes the four critical PM behavioral parameters relevant for engineering the FSI dynamics. Sec.~\ref{sec:Code} formulates the discretized FSI framework used in simulating the dynamics of the fully coupled fluid-PM system. Sec.~\ref{sec:Results} discusses key insights from various fully coupled FSI simulations, with PMs parameterized using the behavioral parameters, to demonstrate their utility in tuning FSI dynamics. Finally, Sec.~\ref{sec:Conclusion} presents a summary and the key takeaways of this article.

\section{Problem configuration and methodology}\label{sec:probmethod}

\begin{figure*}[t!]
  \centering
  \includegraphics[scale=1]{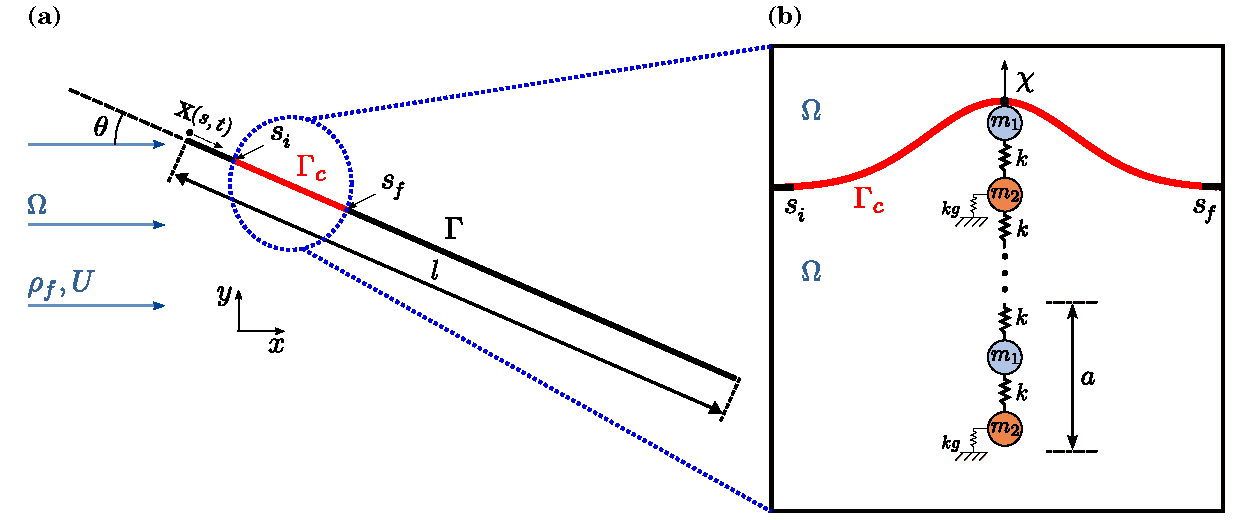}
  \caption{Problem setup and phononic subsystem. 
    (a) Two-dimensional flat plate of chord $l$ at $\theta=12^\circ$ in uniform inflow, showing the fluid domain $\Omega$, the plate surface $\Gamma$, and the compliant section $\Gamma_c\subset\Gamma$ (red).
    (b) Detailed view of the grounded diatomic phononic chain beneath $\Gamma_c$: lumped masses, $m_1$ and $m_2$, inter-mass interaction springs, $k$, and grounding springs, $k_g$, attached to the even masses.}
  \label{fig:schemPM}
\end{figure*}

This section briefly summarizes the FSI configuration considered in this study. A detailed overview of the PM model and the FSI framework that informs the numerical simulations performed in this study is subsequently included in Sec.~\ref{sec:PM} and Sec.~\ref{sec:Code}.

\begin{figure*}[t!]
    \centering
    \includegraphics[width=\textwidth]{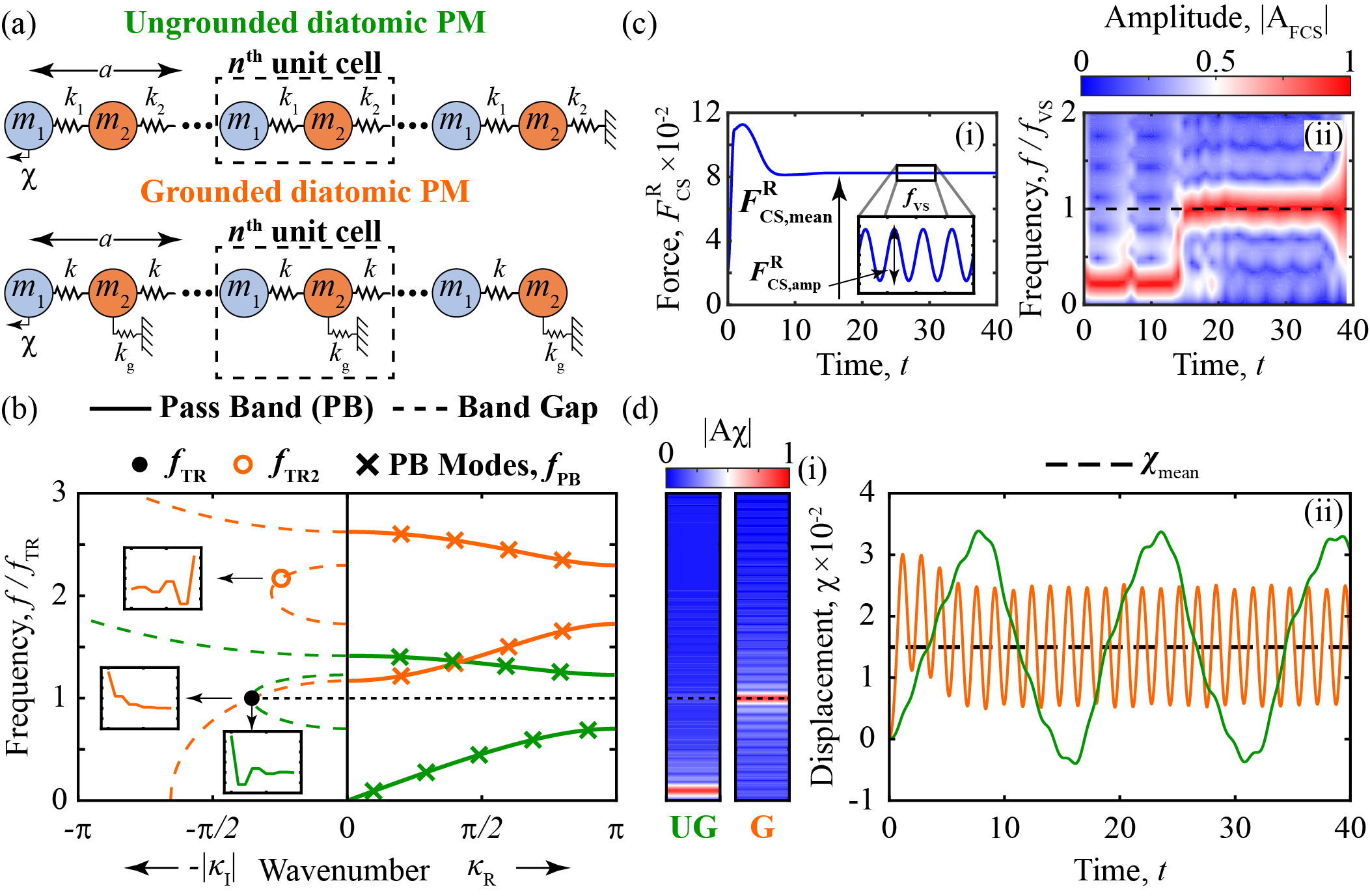}
    \caption{Diatomic Phononic Material. (a) Mass-spring models for an ungrounded and grounded diatomic PMs. (b) Dispersion curves, finite eigenmodes, and truncation resonance eigenmodes (insets) of the ungrounded and grounded diatomic PMs ($N=5$). (c.i) The force signal, and (c.ii) the corresponding frequency spectrogram, of the normal force exerted by the aerodynamic flow onto the CS, $\Gamma_\mathrm{c}$, that will subsequently house the diatomic PM for FSI. (d.i) The frequency spectrum of the diatomic PM responses is plotted as vertical color bars, and (d.ii) the displacement signals of the ungrounded and grounded diatomic PMs when subject to fluid force are depicted in (c.i). Both diatomic PMs feature a primary mode-dominated response. Structural parameters - Ungrounded diatomic PM: $\{m_1,m_2,k_1,k_2\}=\{4.296,13.12,50,50\}$, Grounded diatomic PM: $\{m_1,m_2,k,k_\mathrm{g}\}=\{0.306,0.432,7.04,21.12\}$.}
    \label{fig:UGvsG_DiPM}
\end{figure*}

Our model consists of an incompressible flow ($Re=400$) past a two-dimensional flat plate of chord length $l$ inclined at $\theta=12^\circ$ to the oncoming freestream. These parameters correspond to a configuration where the flow is slightly separated at the trailing edge, with a small degree of unsteadiness. The resulting force signal for this configuration is characterized by a pronounced mean component with small oscillatory fluctuations, reflecting the slight unsteadiness of the flow (see Sec.~\ref{sec:PM} Fig.~\ref{fig:UGvsG_DiPM}). This setup is very near a bifurcation with respect to the angle of attack $\theta$: with a slight decrease in $\theta$, the separation point moves downstream and the flow becomes completely attached and steady. By contrast, at higher values of $\theta$ the separation point moves upstream and induces leading-edge separation and large-scale unsteady vortex shedding. We incorporate a phononic material (PM) into this $\theta=12^\circ$ configuration, to systematically assess what dynamics arise when a PM is incorporated into a flow with clear dynamics that are at a tipping point between amplification towards vortex shedding or pacification towards steadiness. Our aim, in particular, is to identify behavioral parameters that govern distinct features of the modulated flow behavior when the PM is present.

The fluid occupies a rectangular domain $\Omega \subset \mathbb{R}^2$, the plate surface is denoted by $\Gamma$, and a subset $\Gamma_c \subset \Gamma$ is a compliant section (CS) placed at the leading edge, extending 0.1$l$ along the suction side, a location shown to be particularly effective for modulating the flow~\cite{KHAN201779,KANG2020105647,PhysRevFluids.7.024703}. The remainder $\Gamma \setminus \Gamma_c$ is kinematically rigid (see Fig.~\ref{fig:schemPM}a). To focus on the interplay between the aerodynamic flow and the embedded PM, the CS is prescribed kinematically as a Gaussian function whose midpoint displacement, $\chi(t)$, is given by the displacement response of the interface mass, $\chi_1(t)$, of a PM integrated in the CS as a subsurface. This displacement response (and that of the entire PM diatomic chain) is informed by the flow loading on the plate, in a fully (strongly) coupled manner. To facilitate fundamental insights into the effect of the PM on the coupled dynamics, we adopt a simple, lumped mass-spring description of a discrete diatomic PM to model the PM dynamics. Consequently, the diatomic PM~\cite{KushwahaPRL1993,KushwahaIJMPB1996,KhelifSpringer2015} is described by a one-dimensional (1D) chain of two masses interconnected with springs (see Fig.~\ref{fig:schemPM}b for details on the PM model that informs the motion of $\Gamma_c$). The Gaussian shape, dictated by $\chi$, is applied only on $\Gamma_c$; hence, outside $\Gamma_c$ the plate remains rigid. 

Several distinct diatomic PM configurations \cite{BastawrousJASA2022,BaeIJMS2025} could be considered for the proposed study. However, we will demonstrate that only certain types of diatomic PMs are effective for FSI settings featuring aerodynamic lifting bodies~\cite{machado2024fluid}. To establish this point, we first present in Sec.~\ref{sec:PM} the dynamic analysis of two different types of diatomic PMs: an ungrounded and a grounded diatomic PM. In an ungrounded diatomic PM, each mass is only elastically coupled to neighboring masses (i.e., $k_\mathrm{g}=0$ in Fig.~\ref{fig:schemPM}b, except for the final mass where $k_\mathrm{g} = k$). In the grounded diatomic PM, in addition to the inter-mass interaction springs, an additional spring elastically couples every alternate mass to a fixed ground (Fig.~\ref{fig:schemPM}b). In the following section, we demonstrate that these structural differences lead to distinct dynamic behavior in ungrounded and grounded diatomic PMs, which have drastically different consequences for FSI with the given flow.

\section{FSI-relevant Phononic Material behaviors}\label{sec:PM}

This section elucidates several key aspects of phononic material models in the context of FSI. Specifically it: (i) investigates the distinct dynamics found in two different types of diatomic PMs---ungrounded and grounded---and motivates the choice of a grounded diatomic PM for the FSI study; (ii) proposes four FSI-relevant PM behavioral parameters, (iii) relates the behavioral PM parameters to the inherent structural parameters, to provide a map to realize structures with desired behavioral parameters, and (iv) justifies the use of PMs over simpler mass-spring systems. Throughout this section, analysis is performed of the PM diatomic chain in a vacuum, using representative flow forcing from simulations of a fully rigid flat plate (i.e., $\Gamma_c = \{0\}$). The utility of this parametrization will be assessed using fully coupled FSI simulations, in Sec.~\ref{sec:Results}.

\subsection{Dispersion and Eigenmode analysis of Diatomic PMs}\label{subsec:PM_Dispersion}

The general equations of motion of the $n^\mathrm{th}$ unit cell of an ungrounded/grounded diatomic PM, shown in Fig.~\ref{fig:UGvsG_DiPM}a, can be written as:
\begin{align}\label{eq:DiPM_EOM}
    m_1\ddot{\chi}_{2n-1}+k_1(\chi_{2n-1}-\chi_{2n})+&k_2(\chi_{2n-1}-\chi_{2n-2})=0,\\ \nonumber
    m_2\ddot{\chi}_{2n}+k_1(\chi_{2n}-\chi_{2n-1})+&k_2(\chi_{2n}-\chi_{2n+1})+k_\mathrm{g}\chi_{2n}=0,
\end{align}
where $\{m_1,m_2,k_1,k_2,k_\mathrm{g}\}$ represent the two lumped masses, the inter-mass interaction stiffnesses, and the grounding stiffness, respectively. In the ungrounded setting $k_g=0$, and for the grounded case, we take $k_1=k_2=k$ to simplify the mass-spring model, retaining four independent structural parameters. This will be beneficial for developing a one-to-one behavioral-structural parameter map later in Sec.~\ref{subsec:PM_Behaviors}.

For an infinitely periodic diatomic PM, Bloch analysis~\cite{BlochZP1929} on the unit cell (Eq.~\eqref{eq:DiPM_EOM}) provides the dispersion characteristics, i.e., the wavenumber-frequency relations of propagating plane waves within the PM. Therefore, substituting a plane wave \emph{ansatz} of the form:
\begin{equation}
    \bm{\chi}_n(\mathbf{x},t)=[\chi_{2n-1}(t),\chi_{2n}(t)]^\mathrm{T}=\bm{\widetilde{\chi}} e^{i\left(n\kappa-2\pi f t\right)},
    \label{eqn:bloch_ansatz}
\end{equation}
where $\bm{\widetilde{\chi}}$ is the unit cell wave amplitude, $\kappa$, is the non-dimensional wavenumber (normalized by the spatial periodicity, $a$), and $f$ is the wave frequency, into Eq.~\eqref{eq:DiPM_EOM}, and utilizing the Bloch periodicity conditions~\cite{BlochZP1929}, $\bm{\chi_{n\pm1}}=\bm{\chi_n}e^{\pm i \kappa}$, the diatomic PM dispersion relations are calculated as: 
\begin{equation}
    \kappa = -i\log\left(\frac{-b\pm\sqrt{b^2-4}}{2}\right).
    \label{eq:DiPM_Dispersion}
\end{equation}
For the ungrounded PM, $b=\left(k_\mathrm{r}+\frac{1}{k_\mathrm{r}}\right)-\left[1+k_\mathrm{r}-\left(\frac{f}{f_0}\right)^2\right]\left[1+\frac{1}{k_\mathrm{r}}-\frac{m_\mathrm{r}}{k_\mathrm{r}}\left(\frac{f}{f_0}\right)^2\right]$, where $f_0=\frac{1}{2\pi}\sqrt{\frac{k_1}{m_1}}$, and $k_\mathrm{r}=k_2/k_1$ and $m_\mathrm{r}=m_2/m_1$ are the stiffness and mass ratios, respectively. For the grounded PM, $b=2-\left[2-\left(\frac{f}{f_0}\right)^2\right]\left[2+k_\mathrm{gr}-m_\mathrm{r}\left(\frac{f}{f_0}\right)^2\right]$, where $k_\mathrm{gr}=k_\mathrm{g}/k$ is the ground to inter-mass stiffness ratio. Eq.~\eqref{eq:DiPM_Dispersion} characterizes the commensurate number of dispersion branches as the number of internal degrees of freedom within the PM unit cell. Consequently, for the finite ungrounded and grounded diatomic PMs presented here, we have two distinct dispersion branches as illustrated in Fig.~\ref{fig:UGvsG_DiPM}b. The dispersion plots highlight two distinct regimes---(i) the pass bands (solid curves), where each frequency, $f$ is associated with a real wavenumber, $\kappa=\kappa_R$, allowing wave propagation through the entire diatomic PM, and (ii) the band gap (dashed curves), where each frequency, $f$, is associated with a complex wavenumber, $\kappa =i\kappa_I$ or $\pi+i\kappa_I$, forbidding wave propagation in the diatomic PM. That is, for these frequency ranges, the spatial behavior is evanescent, not periodic.

Alternatively, for finite diatomic PMs~\cite{DeymierSpringer2013}, only specific $\kappa-f$ pairs contribute to the PM dynamics. To calculate these wave modes (i.e., structural resonances), we combine Eq.~\eqref{eq:DiPM_EOM} for all unit cells ($n=1,2,\cdots,N$) of a finite diatomic PM to derive the matrix equation of motion: 
\begin{equation}
    \mathbf{M}\ddot{\bm{\chi}}+\mathbf{K}\bm{\chi}=0,
    \label{eq:PM_Matrix_EOM}
\end{equation}
where $\bm{\chi}=[\bm{\chi}_1;\bm{\chi}_2;\cdots;\bm{\chi}_N]$ is the collection of vector displacement of all unit cell masses ($n_\mathrm{d}=2$: no. of unit cell masses), and $\mathbf{M}$ and $\mathbf{K}$, are the mass and stiffness matrices, respectively (see Sec.~\ref{sec:MK_Matrices}). The eigenvalues and the spatial periodicity of eigenvectors, $\bm{\tilde{\chi}}_f$, obtained by solving the eigenvalue problem: 
\begin{equation}
    |\mathbf{K}-4\pi^2 f^2\mathbf{M}|\bm{\tilde{\chi}}_f=0,
    \label{eq:PM_EigProblem}
\end{equation}
correspond to the frequency, $f$, and wavenumber, $\kappa$, of plane waves allowed to propagate in finite PMs, respectively. The $\kappa$ associated with a given $f$ can be identified by performing a spatial Fourier transform of the respective eigenvectors, $\bm{\tilde{\chi}}_f$. These $n_\mathrm{d}\times N=2N$ structural resonances of the finite PMs typically lie on the dispersion branches of their infinite counterparts, with $N$ frequencies lying on each dispersion branch. 

The \scalebox{1.5}{x}-markers in Fig.~\ref{fig:UGvsG_DiPM}b plot the structural resonances corresponding to finite diatomic PMs ($N=5$). Most of the eigenfrequencies of a 5-unit cell finite ungrounded diatomic PM and the grounded diatomic PM lie on the corresponding dispersion pass bands, representing a subset of accessible plane wave solutions from the infinite dispersion super-set. However, a peculiar resonance also exists within the band gap, arising from truncating the infinitely periodic diatomic PM into a finite diatomic PM and from symmetry features of the unit cell. This is commonly referred to as a truncation resonance~\cite{BastawrousJASA2022,HasanPRSA2019,HasanJEL2024} and is denoted by the \scalebox{5}{$.$}-marker (black) in Fig.~\ref{fig:UGvsG_DiPM}b. Note that the grounded diatomic PM also features a second truncation resonance, $f_\mathrm{TR2}\approx2.3f_\mathrm{TR}$ ($6^\mathrm{th}$ mode, \scalebox{1.5}{$\circ$}-marker) within the middle band gap; however, we primarily focus our efforts on engineering the PM behavior with respect to the low-frequency truncation resonance ($1^\mathrm{st}$ mode), as detailed in the following sections. The insets in Fig.~\ref{fig:UGvsG_DiPM}b show the exponentially decaying mode shape of the truncation resonances.

It is worth highlighting that the utility of the truncation resonance and the surrounding phononic band gap in our study are leveraged to intake a broadband fluid forcing and respond at the prescribed truncation resonance frequency with a high energy localization at the fluid-PM interface. This is distinct from prior PM-FSI studies~\cite{hussein2015flow,WilleyJFS23,MichelisPoF2023,BarnesAIAA2021} that primarily leveraged the out-of-phase motion of an ungrounded diatomic PM in an extended frequency region in the phononic band gap around the truncation resonance ($f_\mathrm{TR}<f<f_\mathrm{PB}$) to suppress/amplify TS wave instabilities in wall-bounded flows, that are typically characterized by a dominant single frequency.

\subsection{Grounded diatomic PMs enable meaningful FSI with lifting bodies}
\label{subsec:ground_vs_unground}

We show in Fig.~\ref{fig:UGvsG_DiPM}c.i the load profile, $F_{CS}^\mathrm{R}$, from the flow over a rigid plate, over the portion of the plate $\Gamma_c$ that will be made compliant with the diatomic PM model. (The superscript $R$ denotes that this is a representative force, since that forcing profile will change when the PM is introduced). After a short transient, there are two components to $F_{CS}^\mathrm{R}$: (1) a near-constant, nonzero lift force, $F_\mathrm{CS,mean}$, and (2) an unsteady component due to the time-periodic trailing-edge separation $F_\mathrm{CS,amp}$. We define the frequency of this oscillatory component as $f_\mathrm{VS}$. This frequency will serve as a point of reference within this section and later in assessing how the PM modulates the vortex-shedding behavior from the baseline rigid case. Fig.~\ref{fig:UGvsG_DiPM}c.ii plots the spectrogram of $F_{CS}^\mathrm{R}$ illustrating its multi-frequency composition. Though we expect the fully-coupled FSI dynamics to significantly alter the fluid force ($F_\mathrm{CS}^\mathrm{R}\neq F_\mathrm{CS}$), we can draw important insights, such as the right choice of PM and the relevant PM parameters for FSI, by analyzing the response of a PM in a vacuum to this representative forcing $F_{CS}^\mathrm{R}$.

We present the PM displacement response for the ungrounded and grounded configuration in Fig.~\ref{fig:UGvsG_DiPM}d. Since the focus of this study is aligning a truncation resonance relative to the underlying flow frequency, we will focus on which configuration facilitates more effective interplay between a truncation resonance and the representative flow forcing profile $F_{CS}^\mathrm{R}$. For the ungrounded diatomic PM, the displacement response is dominated by the pass band resonances ($f_\mathrm{PB}$), especially the first mode of the structure, over the higher-order truncation resonance aligned with the vortex shedding frequency, $f_\mathrm{TR}=f_\mathrm{VS}$ (Fig.~\ref{fig:UGvsG_DiPM}d.i). Although a small signature at $f_\mathrm{TR}$ is visible once the unsteady fluid force originates ($F_\mathrm{CS,amp}^\mathrm{R}$), the overall PM dynamics are dominated by the first structural resonance (Fig.~\ref{fig:UGvsG_DiPM}d.ii). This result is intuitive from a structural dynamics perspective: given a broadband input excitation ($F_\mathrm{CS}^\mathrm{R}$), the first structural mode is preferentially excited. Furthermore, in the fully coupled FSI scenario, the presence of viscous effects could induce structural damping at the fluid-PM interface, rendering the first mode even more energetically favorable \cite{HusseinJSV2013}. Therefore, the ungrounded PM can only feature truncation resonances as high-order modes, which are sub-dominant in the structural response compared to the dominant first mode. As such, this configuration is not best suited to effectively interact with the aerodynamic flow considered in this study.

To address this issue, we now consider a grounded diatomic PM, which has its first truncation resonance as its first structural mode. This outcome is because the grounding springs introduce a low-frequency band gap in the PM, from $f=0$ to $f=\min(f_\mathrm{PB})$, that consequently elevates the first pass band to start at a non-zero frequency ($\min(f_\mathrm{PB})$), preventing any rigid body motion \cite{RamakrishnanJSV2025}. This upward shift in the first pass band allows a truncation resonance to exist as a first structural resonance within the low-frequency band gap. Thus, applying a broadband boundary excitation ($F_\mathrm{CS}^\mathrm{R}$) to a grounded diatomic PM will elicit a clean, narrow-band response at the truncation resonance frequency (Fig.~\ref{fig:UGvsG_DiPM}d.i), where the modal response is concentrated at the point of excitation (i.e., fluid-PM interface in case of FSI) due to the localized truncation resonance mode shape (see inset in Fig.~\ref{fig:UGvsG_DiPM}b). Figs.~\ref{fig:UGvsG_DiPM}(d.i-ii) show exactly this: the grounded diatomic PM responds almost purely at the truncation resonance frequency, which is the first structural mode. A supplementary movie (\red{MOV.~1}) of the PM vibration response is also provided for reference.

One may choose alternative forms, e.g.~\cite{BaeIJMS2025}, of a grounded diatomic PM. To facilitate fundamental insights, we focus on a simple, single choice of canonical diatomic PM configuration. We hypothesize that the ensuing dynamics would be qualitatively similar to the results presented in this article, for aligned choices of PM behavioral parameters. However, we leave detailed investigations for a variety of PM configurations to a future study.

\begin{figure*}[t!]
    \centering
    \includegraphics[width=\textwidth]{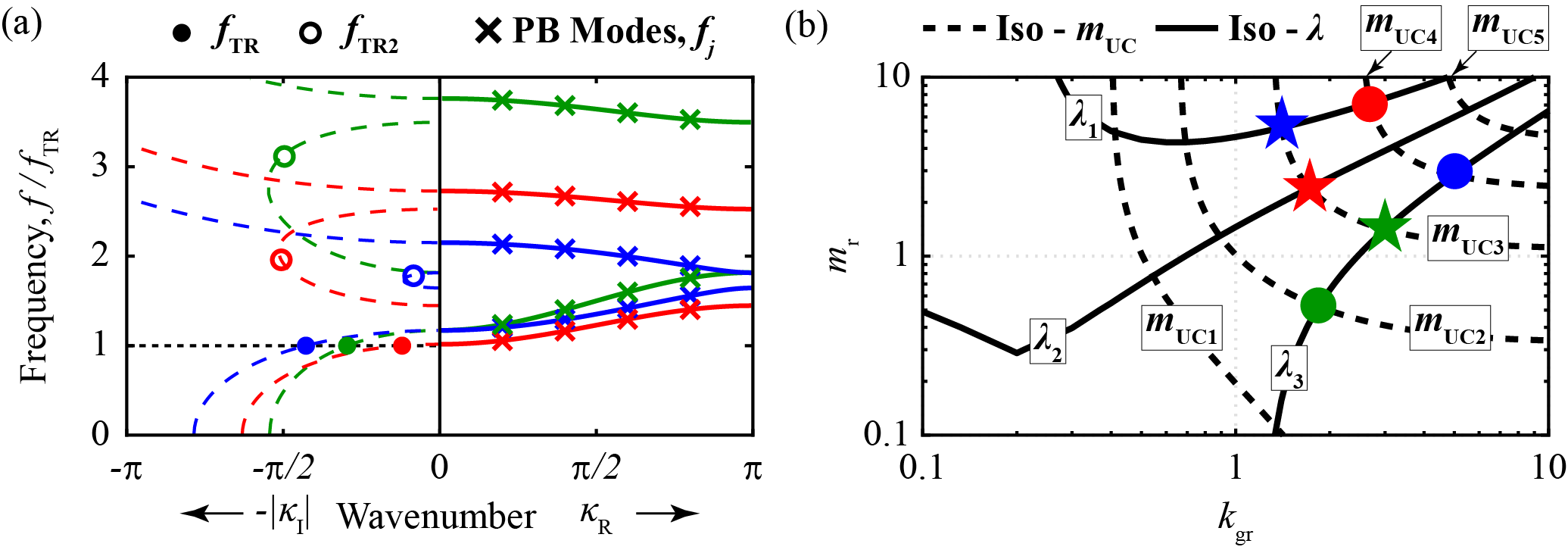}
    \caption{PM behavioral parameters. (a) Dispersion curves of three distinct diatomic PMs engineered with the same truncation resonance frequency, $f_\mathrm{TR}$ (\scalebox{5}{$.$}-marker). The second truncation resonance (\scalebox{1.5}{o}-marker) and pass band modes (\scalebox{1.5}{x}-marker) of the finite diatomic PMs ($N=5$) are also plotted. (b) The behavioral iso-$m_\mathrm{UC}$ and iso-$\lambda$ parameter curves, plotted as a function of the non-dimensional structural parameters - $\{m_\mathrm{r},k_\mathrm{gr}\}$. The structural parameters of the three diatomic PMs explored in (a) are mapped to the \scalebox{5}{$.$}-markers at intersection points - $\{\lambda_3,m_\mathrm{UC2}\}$ (green), $\{\lambda_3,m_\mathrm{UC4}\}$ (blue), and $\{\lambda_1,m_\mathrm{UC4}\}$ (red). The \scalebox{1.5}{$\star$}-markers corresponding to $\{\lambda_{1-3},m_\mathrm{UC3}\}$ will be relevant for discussion in Fig.~\ref{fig:Lambda}.}
    \label{fig:IsoCont}
\end{figure*}

\subsection{Proposed FSI-relevant PM behavioral parameters, and their relation to $\{m_1, m_2, k, k_g\}$}\label{subsec:PM_Behaviors}

This section analyzes the fundamental structural behaviors relevant to the target PM-FSI problem and proposes an alternative PM parameter space in terms of relevant behavioral parameters. This PM parametrization allows for a systematic tuning of distinct static and dynamic behaviors of the FSI system. To facilitate the practical realization of these structural behaviors in computational or experimental PM models, this section also describes a sequential process to calculate the PM structural parameters for a chosen set of target PM behavioral parameters. Though these behavioral parameters are developed in the context of the aerodynamic flow considered in this study, they quantify features of the structural response, such as frequency, dynamic displacement amplitude, and interface deformation, which we argue are relevant for a number of FSI scenarios. Therefore, the behavioral parameters and the subsequent analysis of the parameter map can be adapted according to the specific requirements of distinct FSI studies.

A given set of structural parameters $\{m_1,m_2,k,k_\mathrm{g}\}$ uniquely define the diatomic PM. However, multiple sets of $\{m_1,m_2,k,k_\mathrm{g}\}$ can produce the same $f_\mathrm{TR}$, as shown in Fig.~\ref{fig:IsoCont}a. This redundancy highlights that we cannot simply vary the structural parameters to get distinct FSI responses, since the truncation resonance frequency is expected to strongly influence the FSI. This motivates a broader need to identify alternate parameters that uniquely define the FSI-relevant structural behaviors and allow for systematic tuning of the FSI. Re-parameterizing or normalizing PM parameters in terms of frequencies is relatively common in the structural dynamics community, where parameters are traditionally normalized by frequency quantities such as the band gap edge or Bragg scattering frequency, to generalize the results~\cite{NarisettiJVA2010,ArretcheMSSP2023,RamakrishnanJASA2023}. Additionally, PM design for the majority of applications is focused on the propagating behavior along the structure. By contrast, in this FSI setting, the PM behavior that drives flow modulation is localized near the fluid-PM interface, and the remainder of the chain is only relevant in how it informs dynamics at that interface. Therefore, to our knowledge, the pursuit of a PM parameterization catered to FSI is novel.

We propose four critical PM behavioral parameters that we hypothesize will significantly influence the FSI dynamics: 
\begin{enumerate}
    \item The effective stiffness, $k_\mathrm{eff}$, that governs the mean displacement at the fluid-structure interface where it is subject to a static load (e.g., mean displacement of the interface mass, $\chi_\mathrm{mean}$ in Fig.~\ref{fig:UGvsG_DiPM}d.ii, subject to the near-constant lift force, $F^\mathrm{R}_\mathrm{CS,mean}$, in the given aerodynamic flow setting);
    \item The $\mathrm{J}^\mathrm{th}$ structural resonance frequency, $f_\mathrm{J}$, being engineered; e.g., the truncation resonance frequency, $f_\mathrm{TR}$ ($=f_1$), chosen in this study;
    \item The displacement envelope~\cite{RamakrishnanJSV2025}, $\lambda$, that quantifies the time rate of increase of the displacement amplitude of the interface PM mass, when excited at the chosen structural resonance frequency (graphically represented in Fig.~\ref{fig:Lambda}a.ii);
    \item The total mass of a single PM unit cell, $m_\mathrm{UC}$. This parameter, in combination with $\lambda$, dictates the proximity of other PM structural resonance frequencies to the engineered $f_\mathrm{TR}$ that can influence PM dynamics.
\end{enumerate}

The choice of these behavioral parameters was informed by our own preliminary trial-and-error FSI simulations, where we tested the effect of changing multiple PM behavioral parameters, e.g., effective stiffness, truncation mode shape, band gap width, relative position of the truncation resonance within the band gap, etc., on the fluid dynamics. Prior PM studies considered the effective structural properties~\cite{RamakrishnanJASA2023,NematNasserAIP2011,WojnarPM2014}, such as effective stiffness, primarily to analyze the dynamic response of PMs around structural resonances. Drawing inspiration from these studies, and adapting to the nature of structural loading ($F^\mathrm{R}_\mathrm{CS,mean}$) expected in the given FSI scenario, here we choose the `static' effective stiffness, $k_\mathrm{eff}$, as a relevant behavioral parameter. The choice of the truncation resonance frequency was based on analysis from prior PM-FSI studies~\cite{hussein2015flow,WilleyJFS23,BarnesAIAA2021}, the prospect of engineering this isolated structural resonance, $f_\mathrm{TR}$, in the phononic bandgap relative to the unsteady flow frequency, $f_\mathrm{VS}$, and the subsequent energy localization expected in the PM at the interface around this peculiar resonance as described earlier in Sec.~\ref{subsec:PM_Dispersion}-\ref{subsec:ground_vs_unground}. The displacement envelope, $\lambda$, quantitatively characterizes the envelope of displacement amplitude of a PM at a given structural resonance frequency~\cite{RamakrishnanJSV2025}. This parameter, in particular, is atypical in vibration analysis, which is primarily concerned with characterizing the relative spectral amplitudes of PM masses to determine modal participation factors of the structural resonances of the PM. In addition, this quantity also indirectly captures the effect of the exponential decay rate ($\kappa_I$), i.e., the degree of localization of the truncation resonance mode shape. That said, it was found to be more useful in our trial-and-error simulations than the decay rate directly, our hypothesis for which is that it more directly governs the behavior of the interface mass and the capacity for nonlinear saturation effects when interacting with the flow, motivating its relevance to FSI. Numerical FSI results in Sec.~\ref{sec:Results} will later show the crucial role of $\lambda$ in modulating multiple aspects of the flow behavior. The unit cell mass, $m_\mathrm{UC}$, was chosen as the final behavioral parameter, as it dictates the absolute frequencies of the other structural resonances and their proximity to $f_\mathrm{TR}$ that can influence the spectral composition of the coupled response. It also provides a more natural analog to the structural mass parameter used in a variety of extensively studies FSI systems (see references \cite{riley1988compliant, Gad-el-Hak1996compliant, williamson2004vortex, sarpkaya2004critical, shelley2011flapping} for comprehensive reviews), as it gives a proxy for effective mass not expressible through a single choice of $m_1$ or $m_2$. In addition, $m_\mathrm{UC}$ has a practical relevance in determining the total mass of the PM that needs to be integrated into a lifting body for FSI.

The following equations quantify the proposed PM behavioral parameters in terms of the PM structural parameters, for any general PM model with a mass-spring description:

\begin{equation}
\hat{k}_\mathrm{eff}=\frac{1}{\left\{\hat{\mathbf{K}}^{-1}\right\}_{(1,1)}},
\label{eq:PM_keff}
\end{equation}
\begin{equation}
\hat{f}_\mathrm{J}=\frac{1}{2\pi}\cdot\left\{\mathrm{eig}\left[\hat{\mathbf{M}},\hat{\mathbf{K}}\right]\right\}_\mathrm{(J,J)},
\label{eq:PM_ftr}
\end{equation}
\begin{equation}
\hat{\lambda}=\frac{1}{2(2\pi)^{2n_\mathrm{d}N-1} \hat{f}_\mathrm{J}}\cdot\left\{\frac{\mathrm{adj}\left[\hat{\mathbf{K}}-(2\pi \hat{f}_\mathrm{J})^2\hat{\mathbf{M}}\right]}{\left[\prod_{j\neq \mathrm{J}}(\hat{f}^2-\hat{f}_j^2)\right]\mathrm{det(\hat{\mathbf{M}})}}\right\}_\mathrm{(J,J)},
\label{eq:PM_lambda}    
\end{equation}
\begin{equation}
\hat{m}_\mathrm{UC}=\sum_{j=1}^{n_{d}}\hat{m}_j,
\label{eq:PM_muc}
\end{equation}
\begin{equation}
    k_\mathrm{eff} = \frac{\hat{k}_\mathrm{eff}}{\rho_\mathrm{fluid}U^2},
    m_\mathrm{UC} = \frac{\hat{m}_\mathrm{UC}}{\rho_\mathrm{fluid}hl},
    f_\mathrm{TR} = \frac{\hat{f}_\mathrm{TR}}{U/l},
    \lambda = \frac{\hat{\lambda}}{U},
\label{eq:PM_ND_BParameters}
\end{equation}
where parameters $\hat{\square}$ (Eqns.~\eqref{eq:PM_keff}-\eqref{eq:PM_muc}) and $\square$ (Eq.~\eqref{eq:PM_ND_BParameters}) represent the dimensional and non-dimensional PM behavioral parameters, respectively. $U$ denotes the free stream velocity, $\rho_\mathrm{fluid}$ denotes the 2D fluid density (mass/length$^2$), and, $l$ and $h$, denote the length and thickness of the flat plate. $\mathbf{M}$ and $\mathbf{K}$ (see Sec.~\ref{sec:MK_Matrices}) denote the collated mass and stiffness matrices of the PM, $\{\cdot\}_\mathrm{(J,J)}$ denotes the entry in the $\mathrm{J}^\mathrm{th}$ row and column of the corresponding matrix, and $f_\mathrm{J}$ denotes the $\mathrm{J}^\mathrm{th}$ structural resonance frequency targeted for (mis)alignment with key flow content, e.g., the truncation resonance ($f_1=f_\mathrm{TR}$) in the present study. From hereon, the non-dimensional forms of the PM behavioral parameters, $\{k_\mathrm{eff},f_\mathrm{TR},\lambda,m_\mathrm{UC}\}$, and structural parameters, $\{m_1,m_2,k,k_\mathrm{g}\}$, are considered for PM-FSI simulations.

We anticipate (and subsequently demonstrate in Sec.~\ref{sec:Results}) a significant impact of systematically altering the proposed PM behavioral parameters on equivalent flow characteristics, such as dominant frequencies in quantities such as lift and in the overall vortex-shedding process, making this proposed parametrization suitable for tuning the FSI behavior. Therefore, we argue that this approach of constructing behavioral parameters is foundational even when using more intricate PMs, e.g., auxetic lattices~\cite{ZhangAFM2025}, topological PMs~\cite{VishwakarmaCP2025}, and nonlinear lattices~\cite{PatilAM2022}, for FSI. In those settings, one may require additional behaviors that dictate structural dynamics at the flow-structure interface, or numerically computed variants to (Eqns.~\eqref{eq:PM_keff}-\eqref{eq:PM_muc}) when analytical forms are not available. However, the underlying framework proposed here remains the same.

\subsubsection{Identifying target PM behavioral parameters}\label{subsubsec:PM_Target}

Having proposed FSI-relevant PM behavioral parameters: $\{k_\mathrm{eff},f_\mathrm{TR},\lambda,m_\mathrm{UC}\}$, we must next quantify suitable target behavioral parameters for effective FSI. This will subsequently allow us to derive PM structural parameters using Eqns.~\eqref{eq:PM_keff}-\eqref{eq:PM_muc}, to model the PM dynamics in the fully coupled simulations. This target identification process is demonstrated in the context of the given aerodynamic flow setting. However, one can easily adapt the proposed strategy to identify target parameters for other FSI settings, such as FSI with turbulent channel flows~\cite{LinAIAA2024}.

We perform one-way simulations, with surface motion prescribed over the compliant section (CS), $\Gamma_c$ (i.e., no PM subsurface). These motions are prescribed using $\chi(t)=\chi_\mathrm{mean,ref}+\chi_\mathrm{amp,ref}\sin(2\pi f_\mathrm{ref} t)$. A sweep over the static mean displacement, $\chi_\mathrm{mean,ref}$, dynamic displacement amplitude, $\chi_\mathrm{amp,ref}$, and surface oscillation frequency $f_\mathrm{ref}$ is  performed. Through these, a range of target values is identified to meaningfully impact the flow. Informed by these simulations, we take the following steps:
\begin{enumerate}
    \item The target effective stiffness is calculated as: $k_\mathrm{eff}=F_\mathrm{CS,mean}^\mathrm{R}/\chi_\mathrm{mean,ref}$; i.e., the stiffness is obtained using both the target $\chi_\mathrm{mean,ref}$ and the representative fluid load obtained from the baseline rigid configuration ($F_\mathrm{CS,mean}^\mathrm{R}$). Although the fluid loads in PM-FSI simulations ($F_\mathrm{CS,mean}$) will generally be different than the representative $F_\mathrm{CS,mean}^\mathrm{R}$, we chose $F_\mathrm{CS,mean}^\mathrm{R}$ as a proxy load to avoid a costly iterative process of running several PM-FSI simulations to convergence to the appropriate $k_\mathrm{eff}$. This choice will be validated later in Sec.~\ref{sec:FSI_All} when we discuss the simulation results in detail. The $\chi_\mathrm{mean,ref}$ is chosen from a range of mean surface displacements having a negligible impact on the surrounding aerodynamic flow in the absence of any dynamics (i.e., $\chi_\mathrm{amp,ref}=0$) in the prescribed FSI simulations. Additionally, the choice of $k_\mathrm{eff}$ can impact the customizability range, $\lambda_\mathrm{Range}$, as discussed later in Sec.~\ref{subsec:PM_Benefit}. Therefore, careful consideration is given to choosing the absolute value of $\chi_\mathrm{mean,ref}$, to ensure a broad customizable range for $\lambda$.
    \item The target truncation resonance frequency, $f_\mathrm{TR}$, is chosen such that a commensurate $f_\mathrm{ref}$ produces significant changes in the flow dynamics, in the prescribed simulations.
    \item The target displacement amplitude envelope, $\lambda$, and the target unit cell mass, $m_\mathrm{UC}$, are chosen to systematically explore the behavioral customization range, $\{\lambda_\mathrm{Range},m_\mathrm{UC,Range}\}$, available for the $\{k_\mathrm{eff},f_\mathrm{TR}\}$ chosen in a particular PM-FSI simulation. Furthermore, the $\chi_\mathrm{amp,ref}$ and $f_\mathrm{ref}$ values we found to be useful in the prescribed simulations inform the choice of specific target $\{\lambda,m_\mathrm{UC}\}$ values within the respective customization ranges.
\end{enumerate}

Specifically, for FSI simulations performed in this study, the target PM behavioral parameters are given in Tab.~\ref{tab:PM_BPar}.

\begin{table}[H]
    \centering
    \begin{tabular}{|M{2cm}|M{3cm}|M{3cm}|M{5cm}|}\hline
        $k_\mathrm{eff}$ & $f_\mathrm{TR}$ & $\lambda_{1-3}$ & $m_\mathrm{UC1-5}$\\ \hline
        \multirow{3}*{5.4533} & 0.3128 ($=0.5f_\mathrm{VS}$) & 0.054,0.107,0.161 & 1.39,1.876,2.953,4.759,7.573\\ \cline{2-4}
        & 0.6256 ($=f_\mathrm{VS}$) & 0.107,0.215,0.322 & 0.347,0.469,0.738,1.19,1.893\\ \cline{2-4}
        & 1.2512 ($=2f_\mathrm{VS}$) & 0.215,0.429,0.644 & 0.087,0.117,0.185,0.297,0.473\\ \hline
    \end{tabular}
    \caption{Target PM Behavioral parameters used in fully coupled simulations. ($F_\mathrm{CS,mean}^\mathrm{R}=0.0818$, $\chi_\mathrm{mean,ref}=0.015$, $f_\mathrm{VS}=0.6256$)}
    \label{tab:PM_BPar}
\end{table}

The $f_\mathrm{TR}=0.5f_\mathrm{VS},f_\mathrm{VS},2f_\mathrm{VS}$ are deliberately chosen to study the effect of aligning at a lower, equal, or higher frequency with the $f_\mathrm{VS}$, and to investigate any super/sub-harmonic generation effects in the flow given the $2^\mathrm{nd}$ order nonlinearity of the N-S equation. The $\{\lambda,m_\mathrm{UC}\}$ are chosen to comprehensively span the customization range available for the $\{k_\mathrm{eff},f_\mathrm{TR}\}$ chosen in a particular FSI simulation.

The parameter sweeps of the PM behavioral parameters in Tab.~\ref{tab:PM_BPar} allow us to systematically study the distinct impact of different aspects of the PM behavior on the flow characteristics. Consequently, the above target behavioral parameters are correlated with 45 unique PM structural parameter sets (see Tab.~\ref{tab:FSI_Sims}) that are integrated into the grounded diatomic PM models (Eq.~\eqref{eqn:mass_spring}) in the fully coupled FSI simulations.

\subsubsection{Sequential process to derive PM structural parameters for target PM behavioral parameters}\label{subsubsec:PM_Process}

Having chosen a set of target PM behaviors, this section details a sequential process to calculate the PM structural parameters, by adapting Eqns.~\eqref{eq:PM_keff}--\eqref{eq:PM_muc}, for the grounded diatomic PM. The interaction stiffness, $k$; the interface mass, $m_1$; the non-dimensional mass and grounding stiffness, $\{m_\mathrm{r},k_\mathrm{gr}\}$; are sequentially calculated in that order, to achieve the desired $\{k_\mathrm{eff},f_\mathrm{TR},\lambda,m_\mathrm{UC}\}$.

\begin{enumerate}
    \item First, we create a parameter grid of non-dimensional mass and grounding stiffness ratios, $\bm{\mathcal{G}}:=\{m_\mathrm{r},k_\mathrm{gr}\}\in[0.1,10]\times[0.1,10]$.
    \item From Eq.~\eqref{eq:PM_keff} and the stiffness matrix $\mathbf{K}$ (see Sec.~\ref{sec:MK_Matrices}), we observe that the effective stiffness, $k_\mathrm{eff}$, is only linked to the inter-mass interaction stiffness, $k$, and the grounding stiffness ratio, $k_\mathrm{gr}$. Therefore, given a target $k_\mathrm{eff}$, we calculate the $k$ by solving Eq.~\eqref{eq:PM_keff} for each grid point in $\bm{\mathcal{G}}$, and generate the augmented grid, $\bm{\mathcal{G}^*}:= \{m_\mathrm{r},k_\mathrm{gr}\}_{k}$.
    \item Now, given a target truncation resonance frequency, $f_\mathrm{TR}$, we can calculate the interface mass, $m_1$, by adapting Eq.~\eqref{eq:PM_ftr} for the grounded diatomic PM:
    \begin{equation}
        f_\mathrm{TR}=\frac{1}{2\pi}\sqrt{\frac{k}{m_1}}\cdot\sqrt{\frac{m_\mathrm{r}+k_\mathrm{gr}+1-\sqrt{(m_\mathrm{r}+k_\mathrm{gr}+1)^2-4m_\mathrm{r}k_\mathrm{gr}}}{2m_\mathrm{r}}},
        \label{eq:PM_ftr_GDPM}
    \end{equation}
    and, solving Eq.~\eqref{eq:PM_ftr_GDPM} at each grid point in $\bm{\mathcal{G}^*}$. This generates another augmented grid, $\bm{\mathcal{G}^{**}}:= \{m_\mathrm{r},k_\mathrm{gr}\}_{(k,m_1)}$.
    \item At this stage, two target behavioral parameters, $\{\lambda,m_\mathrm{UC}\}$, remain, that can be one-to-one mapped to each grid point in $\bm{\mathcal{G}^{**}}$, by adapting Eq.~\eqref{eq:PM_lambda} and Eq.~\eqref{eq:PM_muc} for the grounded diatomic PM:
    \begin{equation}
    \lambda=\frac{1}{2\left(2\pi\right)^{4N-1} f_\mathrm{TR}}\cdot\left\{\frac{\mathrm{adj}\left[\mathbf{K}-(2\pi f_\mathrm{TR})^2\mathbf{M}\right]}{\left[\prod_{j=2}^{2N}(f^2-f_j^2)\right]\mathrm{det(\mathbf{M})}}\right\}_{(1,1)},
    \label{eq:PM_lambda_GD}
    \end{equation}
    \begin{equation}
        m_\mathrm{UC}=m_1+m_2=m_1(1+m_\mathrm{r}),
        \label{eq:PM_muc_GD}
    \end{equation}
    and, and solving Eq.~\eqref{eq:PM_lambda_GD} and Eq.~\eqref{eq:PM_muc_GD} at each grid point in $\bm{\mathcal{G}^{**}}$.
\end{enumerate}

For a given choice of four target behavioral parameters: $\{k_\mathrm{eff},f_\mathrm{TR},\lambda,m_\mathrm{UC}\}$, there exists a unique grid point in $\bm{\mathcal{G}^{**}}$ that describes the PM structural parameters: $\{m_\mathrm{r},k_\mathrm{gr}\}_{(k,m_1)}\rightarrow\{m_1,m_2,k,k_\mathrm{g}\}$. The multiple points of intersection in Fig.~\ref{fig:IsoCont}b graphically represent the structural parameters calculated at the end of the above process. The dispersion curves in Fig.~\ref{fig:IsoCont}a can now be traced to the highlighted intersection points (\scalebox{5}{$.$}-markers) in Fig.~\ref{fig:IsoCont}b, giving a clear picture of the identical and unique behavioral features of these different PMs.

\begin{figure*}[t!]
    \centering
    \includegraphics[width=\textwidth]{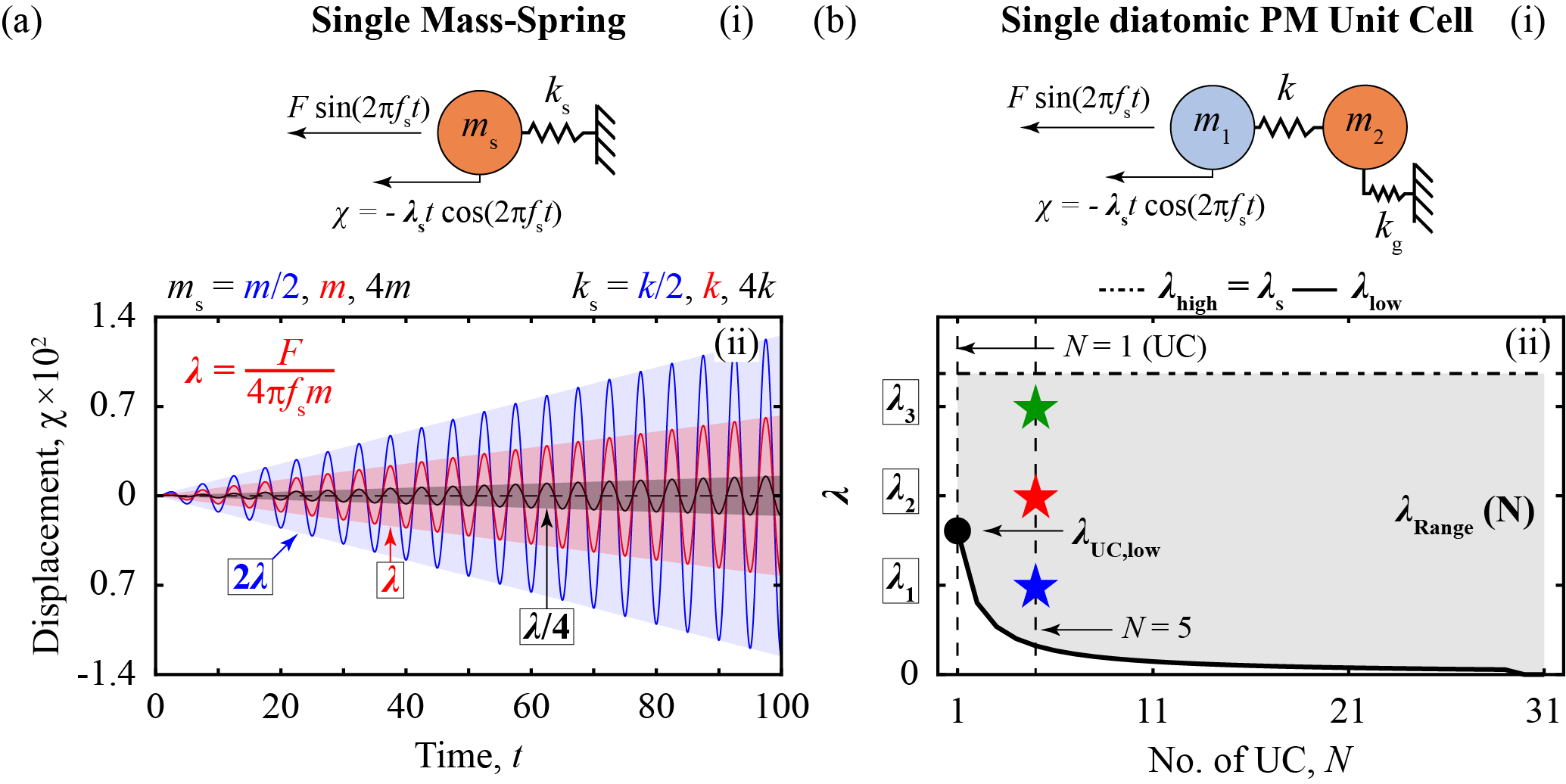}
    \caption{Resonant dynamics and behavioral parameters in different mass-spring models. (a.i) A single mass-spring oscillator subject to resonant force, $F\sin(2\pi f_\mathrm{s}t)$, and (a.ii) displacement response of oscillators with different structural parameters, optimized for different displacement envelopes, $\lambda_\mathrm{s}$, and same resonance frequency, $f_\mathrm{s}$. (b.i) A single unit cell of the grounded diatomic PM depicted in Fig.~\ref{fig:UGvsG_DiPM}a, and (b.ii) the range of displacement envelopes, $\lambda_\mathrm{Range}$, available for customizing the PM behavior when changing the number of unit cells, $N$. The \scalebox{1.5}{$\star$}-markers from Fig.~\ref{fig:IsoCont}b lie within the $\lambda_\mathrm{Range}$ corresponding to PMs ($N=5$ unit cells) considered for FSI simulations in this article. Structural parameters and forcing amplitudes: single mass-spring $\rightarrow\{m_\mathrm{s},k_\mathrm{s},F\}=\{1,1,1\}$, single diatomic PM unit cell: upper bound $\rightarrow\{m_s,k_s,\lambda_\mathrm{high}\}=\{0.353,5.4533,0.36\}$, lower bound $\rightarrow\{m_1,m_2,k,k_\mathrm{g},\lambda_\mathrm{UC,low}\}=\{3.71\times10^{-9},0.739,8.65,11.418,0.17\}$.}
    \label{fig:Lambda}
\end{figure*}

\subsection{Benefit of PM over simpler mass-spring systems}\label{subsec:PM_Benefit}

We now discuss the benefits of the chosen PM system over both a single mass-spring system and a PM diatomic chain with a smaller number of unit cells. For a single mass-spring oscillator parametrized by mass $m$ and stiffness $k$, such as that shown in Fig.~\ref{fig:Lambda}a.i, the presence of only two independent parameters means that only two of the four above behavioral parameters could be tuned independently. For example, one can tune the natural frequency and displacement amplitude of the mass chain as follows. For a chosen natural frequency $f_s = \sqrt{k_s/m_s}$, the displacement amplitude can be tuned by considering the mass-spring system's response to harmonic resonant forcing, $F\sin(2\pi f_\mathrm{s}t)$. The response is of the form, $\chi=-\lambda_\mathrm{s} t\cos{(2\pi f_\mathrm{s}t)}$, indicating an unbounded growth, and the displacement may be tuned by varying the mass $m_s$ (keeping the natural frequency $f_s$ fixed) as: 
\begin{equation}
\lambda_\mathrm{s}=\frac{F}{4\pi f_\mathrm{s}m_s}.
\label{eq:Lambda_SDOF}
\end{equation}

Fig.~\ref{fig:Lambda}a.ii plots the resonant displacement response of three mass-spring systems, $\{m_s,k_s\}=\{m/2,k/2\}$, $\{m,k\},\{4m,4k\}$. These parameters are chosen so that the mass-spring oscillator has a fixed resonance frequency, $f_\mathrm{s}=\sqrt{k/m}$, but distinct $\lambda_\mathrm{s}=\{2\lambda,\lambda,\lambda/4\}$, respectively. Conversely, one could have tuned the spring stiffness $k_s$ to match a target $\chi_\mathrm{mean,ref}$, and subsequently set $m_s$ to match a target natural frequency $f_s$. But then the displacement amplitude $\lambda_s$ will be fully determined by Eq.~\eqref{eq:Lambda_SDOF}. By contrast, the diatomic PM allows for the distinct tuning of an effective stiffness, natural frequency of the first mode (i.e., truncation resonance), the displacement amplitude, and PM mass exposed to the fluid at the interface. The subsequent results section will demonstrate that each of these parameters is relevant to the FSI.

Furthermore, there is a significant benefit to finite-sized diatomic PMs with multiple unit cells over a single unit cell with the same number of free parameters. Specifically, a single diatomic PM unit cell (Fig.\ref{fig:Lambda}b.i) and a finite multi-unit cell diatomic PM (Fig.\ref{fig:UGvsG_DiPM}a) both offer four structural parameters, $\{m_1,m_2,k,k_\mathrm{g}\}$, to define the behavioral parameters, $\{k_\mathrm{eff},f_\mathrm{TR},\lambda,m_\mathrm{UC}\}$. When choosing the structural parameters corresponding to target behavioral parameters - $\{k_\mathrm{eff},f_\mathrm{VS},m_\mathrm{UC3},\lambda_{1-3}\}$ (see Tab.~\ref{tab:FSI_Sims}), we observe a negligible effect of the PM size, $N$ on achieving the desired $\{f_\mathrm{TR},m_\mathrm{UC}\}$. Alternatively, a nominal effect of $N$ on the PM effective stiffness is observed, with the stiffness value saturating to the targeted $k_\mathrm{eff} \ \forall \ N\geq3$. Finally, the effect of $N$ on the PM displacement envelope, $\lambda$, is observed to be the strongest. Fig.~\ref{fig:Lambda}b.ii plots the range of displacement envelopes, $\lambda_\mathrm{Range}$, available for behavioral customizability for different $N$. As the PM size increases, the $\lambda_\mathrm{Range}$ increases, allowing a broader range for tuning the displacement envelope. The upper bound, $\lambda_\mathrm{high}$, is dictated by the displacement envelope, $\lambda_\mathrm{s}$, of a single-spring mass chain with $\{m_\mathrm{s},k_\mathrm{s}\}\rightarrow\{\frac{k_\mathrm{eff}}{4\pi^2f_\mathrm{TR}^2},k_\mathrm{eff}\}$, as this represents a limiting case that localizes all input energy at the interface, while manifesting identical static displacement at the interface as a finite PM. Alternatively, given a sufficient number of unit cells, the lower bound, $\lambda_\mathrm{low}$, can essentially be taken to 0. However, peculiarly for the single unit cell, the lower bound is a non-zero finite value, $\lambda_\mathrm{UC,low}$, that is dictated by the structural parameters.

For a single unit cell, the resonant amplitude envelope can be expressed in the explicit analytical form:
\begin{equation}
    \lambda_\mathrm{UC}=\frac{1}{4\pi f_\mathrm{TR}}\cdot\left\{\frac{1}{m_1+m_2\left[1-\left(\frac{f_\mathrm{TR}}{f_0}\right)^2\right]^2}\right\},
    \label{eq:UC_Lambda}
\end{equation}

A differential analysis of Eq.~\eqref{eq:UC_Lambda}, reveals a lower bound, $\lambda_\mathrm{UC,low}=\frac{1}{4\pi m_\mathrm{UC} f_\mathrm{TR}}$, as $m_1\rightarrow0\Longrightarrow m_2\rightarrow m_\mathrm{UC}, f_0\rightarrow\infty$. The upper bound, $\lambda_\mathrm{UC,high}=\frac{1}{4\pi m_1 f_\mathrm{TR}}$, converges to the single mass-spring behavioral parameters (Eq.~\eqref{eq:Lambda_SDOF}) as $k_\mathrm{g}\rightarrow\infty, f_1\rightarrow f_0 \Longrightarrow f_1=\frac{1}{2\pi}\sqrt{\frac{k}{m_1}}$.

Fig.~\ref{fig:Lambda}b.ii plots the $\lambda_\mathrm{Range}\in[\lambda_\mathrm{low},\lambda_\mathrm{high}]$, as functions of $N$. As indicated in the plot, the lower bound, $\lambda_\mathrm{low}$, asymptotes to 0 for large values of $N$ ($\approx30$), even though the single unit cell for this structural parameter set is bounded by $\lambda_\mathrm{UC,low}>0$. Therefore, finite-sized diatomic PMs provide a higher degree of robustness, accuracy, and customizability in engineering the desired PM behaviors. This benefit is obtained provided sufficiently many unit cells are taken, though the range of design space accessible grows negligibly after $N\approx 5$, which is the value we use in the ensuing work.

\section{FSI Simulation Framework}\label{sec:Code}

We describe here the numerical approach used to incorporate the PM as a subsurface within the compliant section, $\Gamma_c$, of the plate. The numerical methodology leverages the high-fidelity strongly coupled immersed boundary algorithm of \citet{goza2017strongly}. This simulation tool has been validated on a variety of FSI problems, and we leave the reader to reference \cite{goza2017strongly} for full details. We focus instead here on describing how the PM dynamics are incorporated into that simulation framework, with a focus on how the PM subsurface dynamics inform the motion of the compliant section.

\begin{figure*}[t!]
  \centering
  \includegraphics[scale=0.96]{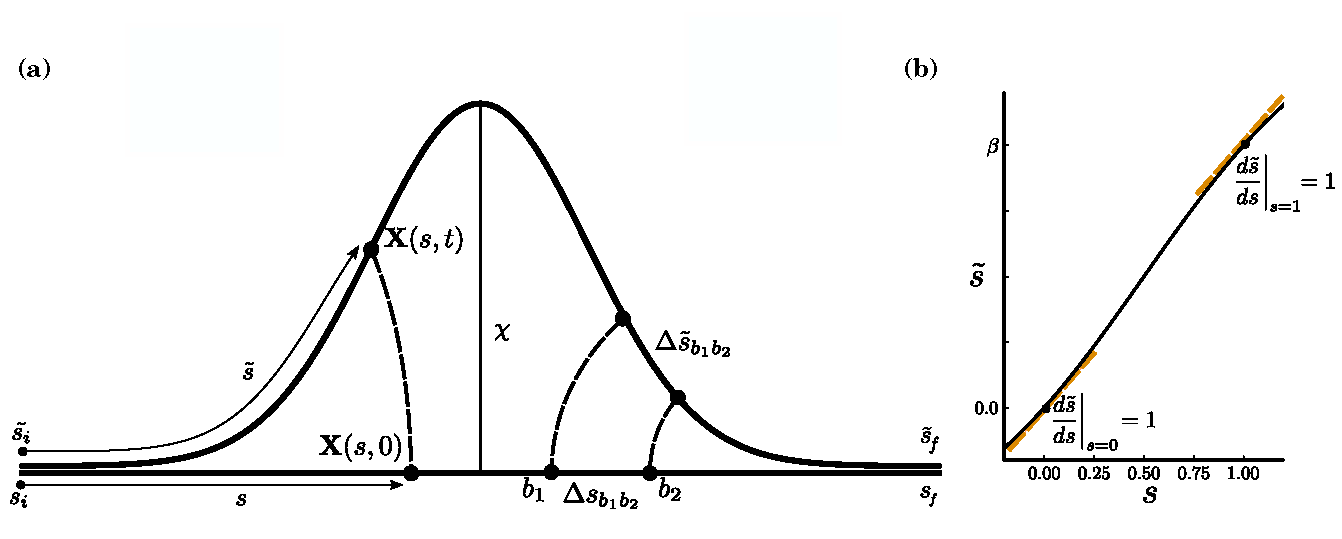}
  \caption{Compliant-section reparametrization and marker redistribution. (a) Undeformed arc-length locations $s\in S=[0,1]$ are mapped to deformed arc length $\tilde s\in\tilde S=[0,\beta(t)]$ by a (b) strictly monotone sigmoid $\tilde s= f(s;a)$ with unit slope at $s=0,1$.}
  \label{fig:compliantsection}
\end{figure*}

To highlight the interplay between the diatomic phononic material and the surrounding aerodynamic flow, and avoid complex coupling between the PM subsurface and the compliant surface, we model the deformation of the compliant section as being fully determined by the PM dynamics. We parametrize the compliant section with respect to a nondimensional arc length variable, $s$, defined as $s=s_i=0$ and $s=s_f=1$ at the leading- and trailing-edge locations of $\Gamma_c$, respectively. We also define the arc length of a material point on $\Gamma_c$ in the deformed configuration as $\tilde{s}$, with leading- and trailing-edge values given by $\tilde{s}_i = s_i = 0$ and $\tilde{s}_f$, respectively. Note that, under this nondimensionalization where $s_f = 1$, the value $\tilde{s}_f$ also represents how much axial stretch the compliant section $\Gamma_c$ has undergone: the axial stretch, $\beta$, is $\beta = \tilde{s}_f/s_f = \tilde{s}_f$. We define the position of a material point on the compliant section in terms of the undeformed arc length, $\mathbf{X}(s,t)$. This position is modeled as having a Gaussian shape, with a peak at the axial center of $\Gamma_c$ defined by the PM displacement at the interface, $\chi := \chi_1$, as
\begin{equation}
\mathbf{X}(s,t)
= \mathbf{X}_{\text{undef}}(s)
+ \chi(t)\,\hat{\mathbf{n}}(s)\, g(s).
\label{eq:CS_disp}
\end{equation} 
In this expression, Eq.~\eqref{eq:CS_disp},
$\mathbf{X}_{\text{undef}}(s)$ is the undeformed plate surface position, $\hat{\mathbf{n}}(s)$ is the unit normal to the undeformed surface at the interface mass, and $g(s) \coloneqq\exp\!\left(-\big[3(2s-1)\big]^{2}/2\right)$ is the unit-peak Gaussian profile set to take on a value of unity at the axial center of $\Gamma_c$, and zero at $s = s_i,$ $s_f$. Thus, outside $\Gamma_c$ the plate remains rigid. Figure \ref{fig:compliantsection}(a) provides a schematic of the compliant section motion, the arc length variables $s$ and $\tilde{s}$, and the deformation of the interface mass $\chi$. Fig.~\ref{fig:compliantsection}(b) provides an illustration of how $\tilde{s}$ is related to $s$, described next.

We construct a model to relate $\tilde{s}$ to $s$, compatible with the definition in Eq.~\eqref{eq:CS_disp}, used to define the Gaussian-shaped surface motion. To facilitate this mapping, we require that (i)  the compliant surface remains undeformed at its endpoints (i.e., $\mathbf{X}(s_i,t)$ and $\mathbf{X}(s_f,t)$ remain unchanged in time); (ii) the deformation is symmetric about the interface mass location (i.e., the axial center of $\Gamma_c$). That is, we require a map $s \rightarrow \tilde{s}$ where the surface points at the ends experience zero axial stretch, while points near the axial center experience the largest stretch. The first physical condition requires that
$\left. \frac{\partial \tilde{s}}{\partial s}\right|_{s_i} = \left. \frac{\partial \tilde{s}}{\partial s}\right|_{s_f} = 1$. To build this intuition, consider two neighboring points (labeled $b_1$ and $b_2$ in Fig.~\ref{fig:compliantsection}a). Denote the arc-length between them in the undeformed configuration by $\Delta s_{b_1b_2}$, and in the deformed configuration by $\Delta \tilde{s}_{b_1b_2}$. The distance between $b_1$ and $b_2$ thus stretches by an amount $\Delta \tilde{s}_{b_1b_2}/\Delta s_{b_1b_2}$, defining the stretch ratio. In the limit as $b_1 \rightarrow b_2$, the stretch ratio is given by the derivative $\frac{d\tilde{s}}{ds}$, which we define as the local (point-wise) stretch. Thus, enforcing no stretch at the left endpoint $s = s_i = \tilde{s}_i$ yields $\left. \frac{d\tilde{s}}{ds}\right|_{s_i} = 1$. A similar argument applies at the right endpoint, leading to $\left. \frac{d\tilde{s}}{ds}\right|_{s_f} = 1$.

A function that satisfies both this derivative requirement, and the need for stretch to be maximal at the axial center, is given by 
\begin{equation}
 \tilde{s}(s)=
  \frac{\beta}{2}\!\left[
    1+\frac{\tanh\!\bigl[\alpha(s-1/2)\bigr]}
             {\tanh(\alpha/2)}
  \right] = \beta f(s;\alpha),
  \label{eq:sigmoidMap}
\end{equation} where $\alpha$ is a width parameter chosen to ensure unit slope at the ends. Note that the PM motion is implicitly built into Eq.~\eqref{eq:sigmoidMap}: the stretch factor $\beta$ is dictated by the Gaussian shape associated with the interface displacement $\chi(t)$. The derivative conditions of $d\tilde{s}/ds = 1$ at the endpoints provide a pathway to determine $\alpha$, via
\begin{equation}
    \frac{df}{ds}(s; \alpha) \;=\; \frac{\alpha}{\tanh\!\bigl(\alpha/2\bigr)}\,\operatorname{sech}^{2}\!\bigl(\alpha(s-\tfrac{1}{2})\bigr),
    \qquad\Rightarrow\qquad
    f'(0;\alpha)=f'(1;\alpha)=\frac{2\alpha}{\sinh \alpha}.
    \label{eq:derivative}
\end{equation}
The rightmost equation in Eq.~\eqref{eq:derivative} may be solved for $\alpha$ at each time instance as the compliant section deforms under the motion of the PM, encoded via deformation of the interface PM mass $\chi$. We emphasize that the axial stretch encoded in $\tilde{s}(s)$ and the Gaussian deformation profile (Eq.~\eqref{eq:CS_disp}) are modeled compatibly in terms of  the same degree of freedom $\chi(t)$, ensuring geometric consistency throughout the compliant region.

The coupled equations are now presented using the immersed boundary formalism of reference \cite{goza2017strongly}. All the quantities are assumed to be dimensionless with lengths normalized by the flat plate chord length $l$, velocities normalized with the freestream velocity $U$, and times by $l/U$. We denote the dimensionless flow velocity and pressure within the fluid domain as $\bm{\mathsf{u}}$ and $p$, respectively, and the surface stress distribution along the plate, which enforces the no-slip condition, as $\bm{\mathsf{f}}$. The dimensionless governing equations are then 
\begin{gather}
\frac{\partial \bm{\mathsf{u}}}{\partial t}+\bm{\mathsf{u}}\cdot\nabla \bm{\mathsf{u}}=-\nabla p +\frac{1}{Re}{\nabla}^2\bm{\mathsf{u}}+\int_{\Gamma}\bm{\mathsf{f}}(\tilde{s},t)\delta(\mathbf{X}(s(\tilde s,t),t)-\bm{\mathsf{x}}) d\tilde{s},
\label{eqn:NS} \\
\nabla\cdot\bm{\mathsf{u}}=0, \label{eqn:continuity} \\
\mathbf{M}\,\frac{d^2\bm{\chi}}{dt^2} + \mathbf{K}\,\bm{\chi}(t) = \bm{\mathbf{f}}_{\text{PM}}(t), \label{eqn:mass_spring} \\
\int_{\Omega}\bm{\mathsf{u}}(\bm{\mathsf{x}}, t)\delta(\bm{\mathsf{x}}-\mathbf{X}(s(\tilde s,t),t))d\bm{\mathsf{x}}=\frac{d\,\mathbf{X}(s(\tilde s,t),t)}{dt} \quad \forall \mathbf{X}\in \Gamma. \label{eqn:no-slip}
\end{gather}
Eq.~\eqref{eqn:NS} describes the momentum equations for the flow, incorporating the effects of stresses from the flat plate surface to enforce the no-slip boundary condition, including the contribution from the compliant section $\Gamma_c$. Eq.~\eqref{eqn:continuity} is the conservation of mass for incompressible flow. Eq.~\eqref{eqn:mass_spring} specifies the matrix equations of motion for the diatomic PM. Here, the loading vector
\begin{equation}
    \bm{\mathbf{f}}_{\text{PM}}(t)=\big[F_\mathrm{CS}(t),\,0,\,\ldots,\,0.\big]^\top,
    \label{eq:f_pm}
\end{equation} captures the fluid forces integrated along the CS in its first entry; all other entries are zero as they pertain to parts of the PM that are not influenced directly by the external flow. That is, the flow load applied to the structure is at the interface location alone, and this stimulus drives PM motion which in turn modulates the flow. The loading applied to the diatomic PM chain is computed as the integral of the normal component of the surface stress $\bm{\mathsf{f}}$ along $\Gamma_c$, given by
\begin{equation}
    F_\mathrm{CS}(t) 
    = \int_{\Gamma_c} 
      \bm{\mathsf{f}}(\tilde{s},t)\,\cdot\,\hat{\mathbf{n}}(\tilde{s})\, d\tilde{s}.
    \label{eq:F_Gamma_c}
\end{equation}
Finally, Eq.~\eqref{eqn:no-slip} enforces the no-slip boundary condition, ensuring that the velocity of the flow and the plate match at every material point along the plate, including over the compliant section $\Gamma_c$. 

Note that in both the immersed boundary source term in Eq.~\eqref{eqn:NS} and in the no-slip condition in Eq.~\eqref{eqn:no-slip}, surface quantities are evaluated with respect to the deformed arc-length $\tilde{s}$, with mappings to the undeformed arc length $s$ utilized as needed, from Eq.~\eqref{eq:sigmoidMap} and its unique inverse. This approach has two benefits. (i) It allows the integral in Eq.~\eqref{eqn:NS} to be written without a Jacobian mapping from the deformed to undeformed state. (ii) The dynamical system (Eqns.~\eqref{eqn:NS}--\eqref{eqn:no-slip}) is time advanced after spatial discretization, described next. The immersed surface, including the sub-portion $\Gamma_c$, is represented using a set of discrete points after this discretization process. Working in terms of $\tilde{s}$ allows for these points to be dynamically updated throughout the simulation, to maintain a uniform spacing as the compliant section deforms under its Gaussian shape. This outcome avoids points becoming either too far apart---resulting in an effectively porous surface---or becoming too closely spaced---leading to ill-conditioning issues in the time advancement process \cite{goza2016accurate}.

Eqns.~\eqref{eqn:NS}--\eqref{eqn:continuity} are discretized in space using standard central second-order finite difference stencils, and recast in a discrete vorticity-streamfunction framework as in \citet{colonius2008fast}. Conditions on the outer boundaries of the flow domain are treated using a multi-domain approach, as described in reference \cite{colonius2008fast}. The final form of the spatially discrete equations, using an overdot to denote differentiation with respect to time, is
\begin{gather}
    \mdisc{C}^{\!\top}\mdisc{C}\,\dot{\mdisc{s}}
= -\,\mdisc{N}\!\big(\mdisc{C}\,\mdisc{s}\big)
  + \mdisc{C}^{\!\top}\mdisc{L}\,\mdisc{C}\,\mdisc{s}
  - \mdisc{C}^{\!\top}\mdisc{E}^{\!\top}\mdisc{f},
\label{eqn:NS_sd} \\
\mathbf{M}\,\dot{\bm{\sdisc{\xi}}} + \mathbf{K}\,\bm{\chi}
= \widetilde{\mdisc{Y}}\,\mdisc{W}(\bm{\chi})\,\mdisc{f}\;=\;\mdisc{f}_{\mathrm{PM}},\label{eqn:mass_spring_sd} \\
\dot{\bm{\chi}} = \bm{\sdisc{\xi}},\label{eqn:identity_sd} \\
\mdisc{E}\,\mdisc{C}\,\mdisc{s} - \mdisc{Y}\,\bm{\sdisc{\xi}} = \mathbf{0}. \label{eqn:no-slip_sd}
\end{gather}
Eq.~\eqref{eqn:NS_sd} is the spatially discrete analog of Eq.~\eqref{eqn:NS}, where discrete curl operators $\mdisc{C}$, $\mdisc{C}^\top$ are used to cast the governing momentum equation in terms of a discrete streamfunction, $\mdisc{s}$. This approach formally removes the pressure term from the momentum equation, and removes the continuity equation altogether, because defining the velocity $\mdisc{u}$ in terms of the streamfunction via $\mdisc{u} = \mdisc{C\,s}$ automatically satisfies the divergence free condition. The quantity $\mdisc{f}$ is the spatially discrete counterpart of the space-time continuous surface stress, $\bm{\mathsf{f}}$ used to impose the no-slip condition. $\mdisc{N}(\cdot)$ and $\mdisc{L}$ denote the discretization of the nonlinear advective and viscous Laplacian terms, respectively. The operator $\mdisc{E}^\top$ is the spatially discretized counterpart of the integral operator with the delta kernel from the momentum equation Eq.~\eqref{eqn:NS}.

The equations governing the solid are modeled as a mass-spring system. Since this model is already space-discrete it is left unchanged from its prior form Eq.~\eqref{eqn:mass_spring}, except that it is written in first-order form for consistency with the flow governing equations, with the surface velocity $ \bm{\sdisc{\xi}} =\dot{\bm{\chi}}$ defined as a distinct state variable. The term $\mdisc{W}$ is a weighting matrix arising through spatial discretization to ensure that the surface stresses applied to the structural equations of motion are accurate, and avoid the spurious oscillations provided by many immersed boundary methods \cite{goza2016accurate}. The load-assembly operator $\widetilde{\mdisc{Y}}$ is the spatial discretization of Eqns.~\eqref{eq:f_pm}--\eqref{eq:F_Gamma_c}, integrating the normal component of the surface stress along $\Gamma_c$ via quadrature. 

Eq.~\eqref{eqn:no-slip_sd} is the spatially discrete variant of Eq.~\eqref{eqn:no-slip}, where $\mdisc{E}$ contains the spatially discrete analog of the integral operator with the delta kernel. Note that, as demonstrated in \citet{taira2007immersed}, this discrete integral operator can be cast as the transpose of $\mdisc{E}^\top$, provided a straightforward scaling is incorporated into the discrete surface stress $\mathbf{f}$. $\mdisc{Y} $ is the kinematic counterpart operator, which is used in the semi-discrete equation for the no-slip condition and originates from the center of the velocity of the CS by distributing this velocity across a Gaussian profile and accounts for the zero velocity of the rigid body parts. Detailed constructions (with indices and sparsity) of $\mdisc{Y}$ and $\widetilde{\mdisc{Y}}$ operators are provided in Sec. \ref{sec:OperatorY_&_Y_tilde}

Eq.~\eqref{eqn:NS_sd} is discretized in time with an Adams-Bashforth scheme for the nonlinear terms, a trapezoid method for the diffusive term, and an implicit treatment of the surface stress term, ensuring that the no-slip condition (Eq.~\eqref{eqn:no-slip_sd}) is enforced to within machine precision at each time step, including the current one. The dynamics of the diatomic PM (Eqns.~\eqref{eqn:mass_spring_sd}--\eqref{eqn:identity_sd}) are advanced in time using a second-order implicit Newmark scheme. This strongly coupled fluid-structure interaction (FSI) approach results in a nonlinear algebraic system of equations, which is solved using Newton’s method. A block-LU factorization is applied to the linearized system of equations for an efficient iterative solution process \cite{goza2017strongly}. The simulation
parameters and domain independence results are reported in Sec. \ref{sec:grid_convergence}.

\section{Results of FSI Simulations}\label{sec:Results}

\begin{figure*}[t!]
    \centering
    \includegraphics[width=\textwidth]{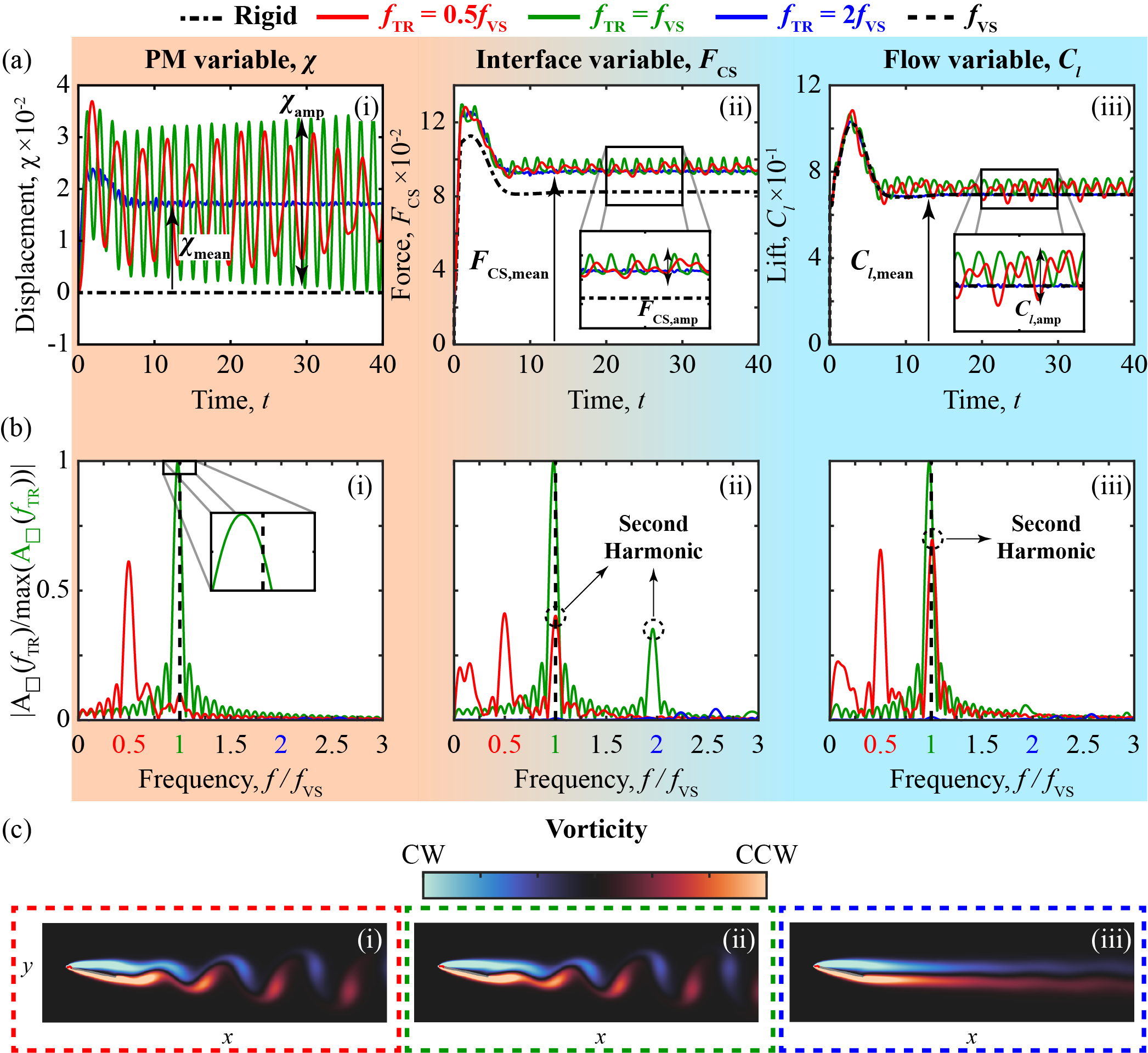}
    \caption{Vortex-shedding process as a function of different PM truncation resonance frequencies. (a.i) The surface displacement, $\chi$ (PM variable), (a.ii) interface fluid force, $F_\mathrm{CS}$ (interface variable), and (a.iii) lift force, $C_{l}$ (flow variable) signals obtained from FSI simulations highlighted in Fig.~\ref{fig:FFT_CBars}. (b.i-iii) The Fourier frequency spectrum of the post-transient portion ($t\geq20$) of the respective signals. (c) Vorticity maps of the FSI domains at $t=40$, for PMs with: (i) $f_\mathrm{TR}=0.5f_\mathrm{VS}$, (ii) $f_\mathrm{TR}=f_\mathrm{VS}$, and (iii) $f_\mathrm{TR}=2f_\mathrm{VS}$ (CW - Clockwise, CCW - Counter Clockwise). Structural parameters: $f_\mathrm{TR}=0.5f_\mathrm{VS}\rightarrow\{m_1,m_2,k,k_\mathrm{g}\}=\{0.866,2.086,8.02,13.794\}$, $f_\mathrm{TR}=f_\mathrm{VS}\rightarrow\{m_1,m_2,k,k_\mathrm{g}\}=\{0.315,1.577,6.1,48.617\}$, and $f_\mathrm{TR}=2f_\mathrm{VS}\rightarrow\{m_1,m_2,k,k_\mathrm{g}\}=\{0.054,0.13,8.02,13.794\}$.}
    \label{fig:Rigid_vs_PM}
\end{figure*}

A total of 45 unique PMs were chosen to systematically vary the FSI dynamics. To focus on the dynamic interaction, the effective stiffness is kept constant at $k_\mathrm{eff}=5.4533$, to target a mean displacement of $\chi_\mathrm{mean,ref}=0.015$. The remaining three behavioral structural parameters are varied in a parameter matrix (Tab.~\ref{tab:FSI_Sims}), to quantify their effect on the FSI dynamics: 3 truncation resonance frequencies ($f_\mathrm{TR}=0.5f_\mathrm{VS},f_\mathrm{VS},2f_\mathrm{VS}$), 3 displacement envelopes ($\lambda_{1-3}$), and 5 unit cell masses ($m_\mathrm{UC1-5}$). Note that the displacement envelopes and unit cell masses are distinct for PMs with different truncation resonance frequencies. However, $\{m_\mathrm{UC1-5},\lambda_{1-3}\}$ are chosen such that they are similarly spaced across the respective $\{m_\mathrm{UC,Range},\lambda_{Range}\}$ behavioral customization range available for the chosen $\{k_\mathrm{eff},f_\mathrm{TR}\}$ (see Sec.~\ref{subsubsec:PM_Target} and Tab.~\ref{tab:FSI_Sims}).

For a given $f_\mathrm{TR}$, the 15 intersection points of the iso-$\lambda$ and the iso-$m_\mathrm{UC}$ curves in Fig.~\ref{fig:IsoCont}b define the PM structural parameters. The numerical values of the PM behavioral and structural parameters used in the FSI simulations are included in the appendix (Tab.~\ref{tab:FSI_Sims}).

\subsection{Representative FSI results when varying $f_\mathrm{TR}$}\label{subsec:FSI_fTR}

We first give an overview of representative FSI simulations to understand the general dependence of the FSI on PM truncation resonance frequency, when it is either aligned or misaligned with the inherent fluid vortex frequency, $f_\mathrm{VS}$. Fig.~\ref{fig:Rigid_vs_PM}a compares three key dynamical quantities for three representative PMs with $f_\mathrm{TR}=0.5f_\mathrm{VS},f_\mathrm{VS},2f_\mathrm{VS}$: the (i) surface displacement at the center of the compliant section (CS), $\chi$; (ii) the interface force, $F_\mathrm{CS}$, constituting the integral of the normal stress along the CS $\Gamma_c$, of the plate; (iii) the lift force, $C_{l}$ obtained by integrating the vertical component of surface stress along the entire plate $\Gamma$. Note that all three frequency spectra (i.e, $\mathrm{A}_\Box(f_\mathrm{TR}=0.5f_\mathrm{VS},f_\mathrm{VS},2f_\mathrm{VS})$, where $\Box=
\chi,F_\mathrm{CS},C_{l}$) in Fig.~\ref{fig:Rigid_vs_PM}b are normalized by the maximum Fourier amplitude, $\mathrm{max}( \mathrm{A}_\Box(f_\mathrm{TR}=f_\mathrm{VS}))$. This Fourier amplitude scaling allows us to study and compare the relative mechanical energy distribution in the three representative FSI configurations.

All three PMs exhibit a similar $F_\mathrm{CS,mean}$ and $\chi_\mathrm{mean}$, but different dynamics in response to the unsteady fluid force, $F_\mathrm{CS,amp}$. The strongly similar mean displacements are a result of the same static stiffness, $k_\mathrm{eff}$, for the three PMs. This outcome justifies the use of $k_\mathrm{eff}$ as one behavioral parameter that drives the static system response. The slight change in mean displacement in the FSI simulations compared to the target displacement ($\chi_\mathrm{mean,ref}$) means the mean interface force in the post-transient state, $F_\mathrm{CS,mean}=k_\mathrm{eff}\chi_\mathrm{mean}$, is also slightly altered from the rigid case. We show in Sec.~\ref{subsec:FSI_S&D} that, across all simulated cases, this difference has a negligible effect on the FSI dynamics, enabling us to distinguish the specific impact of static behavior tuned by $k_{\text{eff}}$ versus dynamic FSI behavior induced by the PM.

The PMs with $f_\mathrm{TR}=0.5f_\mathrm{VS},f_\mathrm{VS}$ in Fig.~\ref{fig:Rigid_vs_PM}(a,b.i) yield stronger FSI effects than for $f_\mathrm{TR}=2f_\mathrm{VS}$. In both the $f_\mathrm{TR}=0.5f_\mathrm{VS}$ and $f_\mathrm{TR}=f_\mathrm{VS}$ scenarios, there are significant dynamic displacements of the PM ($\chi_\mathrm{amp}$) concentrated at a frequency very near the truncation resonance frequency, $f_\mathrm{TR}$. We hypothesize that the slight deviation from $f_{TR}$ is due to a non-linear effect arising from the fluid-PM coupling (see Sec.~\ref{subsubsec:NL_FSI} for a brief discussion). Nevertheless, the subsequent effect of PM dynamics on the flow is distinct for $f_\mathrm{TR}=0.5f_\mathrm{VS}$ and $f_\mathrm{TR}=f_\mathrm{VS}$. When the dominant temporal frequencies excited within the PM interact with the aerodynamic flow, the quadratic flow nonlinearity leads to second harmonic generation at the interface, as seen in the interface force $F_{CS}$ in Fig.~\ref{fig:Rigid_vs_PM}b.ii, with visible spectral peaks at $f=2f_\mathrm{TR}$. The spectral peaks at $f=f_\mathrm{TR}$ persist in both the $F_\mathrm{CS}$ and $C_{l}$ signals, as a result of the resonant dynamics of the PM; however, the second harmonic frequencies, $f=2f_\mathrm{TR}$, generated at the interface (manifested in $F_\mathrm{CS}$) only persist beyond the CS and along the entire plate (manifested in $C_{l}$) if they lie in the vicinity of the vortex shedding frequency, $f_\mathrm{VS}$. Therefore, as the flow evolves, only the second harmonic, $f=f_\mathrm{VS}$, generated in the $f_\mathrm{TR}=0.5f_\mathrm{VS}$ case survives within the flow, whereas the second harmonic $f=2f_\mathrm{VS}$ does not persist in $C_{l}$ for the $f_\mathrm{TR}=f_\mathrm{VS}$ case (Fig.~\ref{fig:Rigid_vs_PM}b.iii). These dynamics are consistent with the time-trace history of the $F_{CS}$ and $C_{l}$ signals for the $f_\mathrm{TR}=0.5f_{VS}$ and $f_\mathrm{TR}=f_{VS}$ cases in Fig.~\ref{fig:Rigid_vs_PM}(a.ii-iii). 

Fig.~\ref{fig:Rigid_vs_PM}(c.i-iii) captures a snapshot of the vortex-shedding process in the simulation domain at $t=40$, for each of these PM-FSI configurations. The vorticity field for $f_\mathrm{TR}=2f_\mathrm{VS}$ (Fig.~\ref{fig:Rigid_vs_PM}c.iii) is visually similar to that of the rigid inclined flat plate (not pictured), which corroborates the relatively muted dynamics seen in the time domain plots. Alternatively, a clear vortex-shedding process is visible in the wake for PMs with $f_\mathrm{TR}=0.5f_\mathrm{VS},f_\mathrm{VS}$. The distinct dynamics for $F_\mathrm{CS}$, $C_{l}$ are associated with different time scales over which the vortical features are generated. This manifests as vortices with visually different spatial behavior. A supplementary movie \red{(MOV.~2)} of the vortex-shedding process for the three FSI configurations described in Fig.~\ref{fig:Rigid_vs_PM} is also provided for reference.

\subsection{Frequency spectrum analysis of the FSI system varying $\{f_\mathrm{TR},\lambda,m_\mathrm{UC}\}$}\label{sec:FSI_All}

\begin{figure*}[t!]
    \centering
    \includegraphics[width=\textwidth]{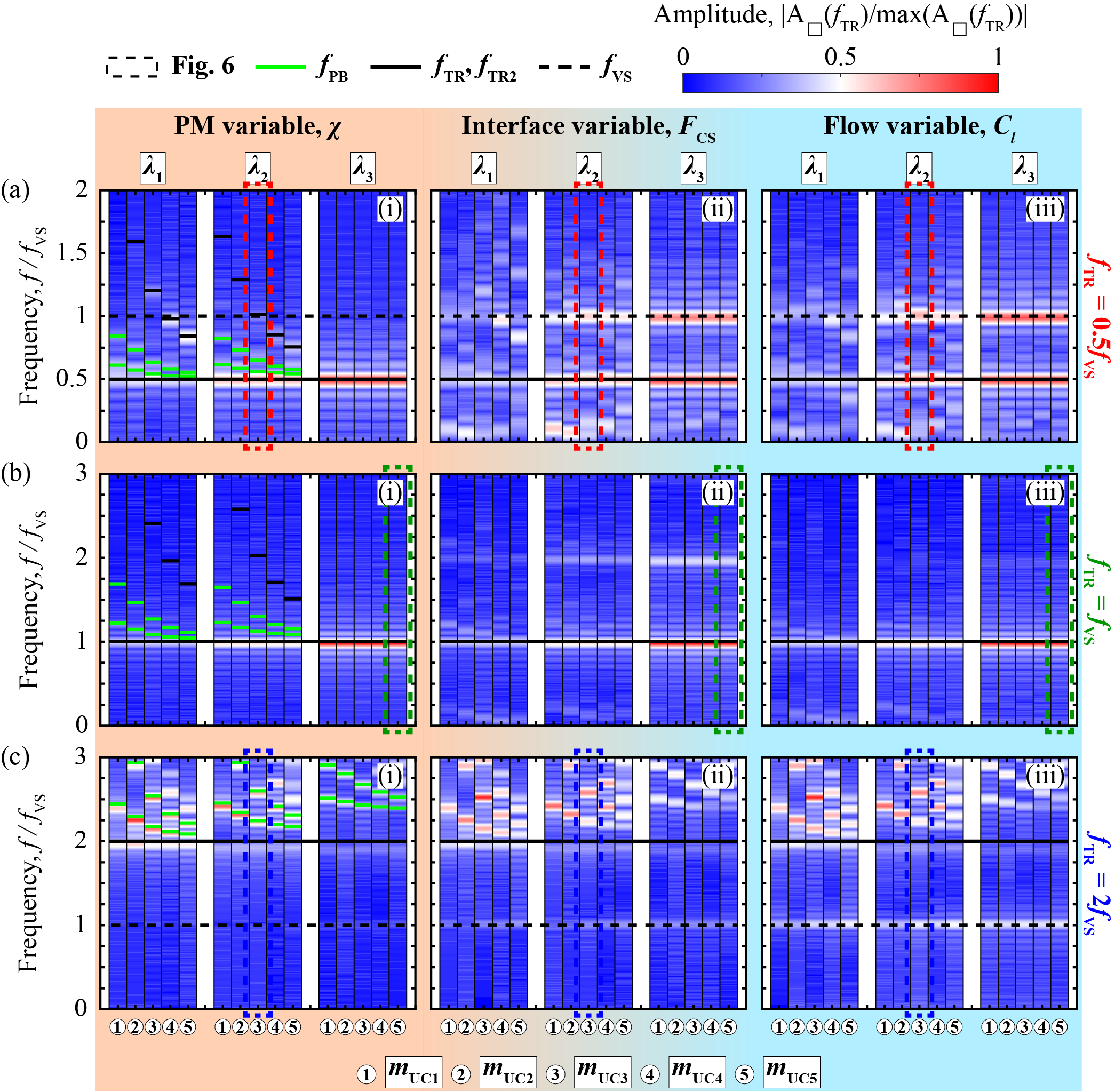}
    \caption{Fourier frequency spectra in the post-transient state ($t\geq20$) in terms of: (i) surface displacement, $\chi$ (PM variable), (ii) interface fluid force, $F_\mathrm{CS}$ (interface variable), and (iii) lift force, $C_{l}$ (flow variable) signals from all 45 FSI configurations explored in this article. FSI configurations featuring PMs with (a) $f_\mathrm{TR}=0.5f_\mathrm{VS}$, (b) $f_\mathrm{TR}=f_\mathrm{VS}$, and (c) $f_\mathrm{TR}=2f_\mathrm{VS}$. Structural parameters are given in Table~\ref{tab:FSI_Sims}.}
    \label{fig:FFT_CBars}
\end{figure*}

This section presents a comprehensive analysis of the frequency spectrum, the coupling strength of fluid-PM dynamics, and nonlinear phenomena observed in the FSI simulations. Fig.~\ref{fig:FFT_CBars} summarizes the frequency spectrum data obtained from all 45 FSI configurations explored in this article. Each row in Fig.~\ref{fig:FFT_CBars}(a-c) plots the frequency spectra corresponding to the 15 FSI configurations explored for a given $f_\mathrm{TR}$, while each column plots the post-transient frequency spectra of the (i) PM variable (surface displacement, $\chi$), (ii) interface variable (interface force, $F_\mathrm{CS}$), and (iii) flow variable (lift force, $C_{l}$) for the different values of $f_\mathrm{TR}$. Note that we subtract the mean quantities, i.e., $\chi_\mathrm{mean}$, $F_\mathrm{CS,mean}$, and $C_{l,\mathrm{mean}}$, from their respective signals to filter out the low-frequency components and accurately capture the Fourier frequency spectra corresponding to the post-transient FSI dynamics. Within each figure, we have two hierarchies of columns, one showing changes due to varying $m_\mathrm{UC1-5}$ and the other showing changes due to varying the prescribed $\lambda_{1-3}$. The parameters studied in Fig.~\ref{fig:Rigid_vs_PM} are highlighted by dashed boxes in Fig.~\ref{fig:FFT_CBars}. In addition to the inferences drawn from Fig.~\ref{fig:Rigid_vs_PM}, collectively analyzing all simulation results in Fig.~\ref{fig:FFT_CBars} reveals more details about the emerging FSI dynamics when systematically varying the PM behavioral parameters $\{f_\mathrm{TR},\lambda,m_\mathrm{UC}\}$.

Note that all the frequency spectra at a given $f_\mathrm{TR}$  in Fig.~\ref{fig:FFT_CBars} are normalized by the maximum Fourier amplitude at that $f_\mathrm{TR}$ (i.e., each $\mathrm{A}_\Box(f_\mathrm{TR})$, where $\Box=\chi,F_\mathrm{CS},C_{l}$,  is normalized by $\mathrm{max}( \mathrm{A}_\Box(f_\mathrm{TR})$). Consequently, the narrow-band red spectral peaks highlight the strongest fluid-PM coupling at a given $f_\mathrm{TR}$. In addition to allowing us to characterize the distinct frequency spectra observed, this Fourier amplitude scaling allows us to study the relative mechanical energy distribution in the PMs and the flow in each FSI configuration.

\subsubsection{FSI Dynamics: $f_\mathrm{TR}=0.5f_\mathrm{VS}$}

Consider the PMs featuring a truncation resonance at $f_\mathrm{TR}=0.5f_\mathrm{VS}$, i.e., Fig.~\ref{fig:FFT_CBars}a. For a low-displacement envelope, $\lambda_1$, the PM exhibits a broadband frequency response. A closer look at the frequency spectrum of the surface displacement reveals a relatively significant excitation of the pass band frequencies, $f_\mathrm{PB}$ (green solid lines in Fig.~\ref{fig:FFT_CBars}a.i show the first two pass band eigenfrequencies), in addition to the first eigenmode, $f_\mathrm{TR}$. Consequently, $m_\mathrm{UC}$ dictates the frequency spectrum at low displacement envelopes (since $k_\mathrm{eff}$, $f_\mathrm{TR}$, and $\lambda$ are all fixed), as the unit cell mass essentially determines the location and magnitude of the pass band frequencies. Commensurate frequencies and their second harmonics get excited in the flow at the interface, leading to a more broadband fluid dynamics at the fluid-PM interface as seen in the $F_\mathrm{CS}$ spectrum in Fig.~\ref{fig:FFT_CBars}a.ii. However, only frequencies in the vicinity of the primary harmonics excited in the PM, i.e., $f_\mathrm{TR}$ and $f_\mathrm{PB}$, and the global attractor, $f_\mathrm{VS}$, persist in the flow as seen in the $C_{l}$ spectrum in Fig.~\ref{fig:FFT_CBars}a.iii.

Results also show that these pass band frequencies approach each other and decrease in magnitude as $m_\mathrm{UC}$ increases ($m_\mathrm{UC1}\rightarrow m_\mathrm{UC5}$), consistent with the frequency being inversely proportional to mass ($f\propto 1/\sqrt{m}$). Another peculiar component of the frequency spectrum is the second truncation resonance, $f_\mathrm{TR2}$, lying within the $2^\mathrm{nd}$ band gap (between the diatomic PM pass bands; c.f., Fig.~\ref{fig:UGvsG_DiPM}b). As the $f_\mathrm{TR2}$ approaches the global attractor frequency, $f_\mathrm{VS}$, we see a significant peak in $\chi$, $F_{CS}$, and $C_{l}$ at this frequency. In fact, for PMs with unit cell mass $m_\mathrm{UC4,5}$, the mechanical energy concentration at $f_\mathrm{TR2}$ is larger than the pass band components, $f_\mathrm{PB}$, and the primary truncation resonance, $f_\mathrm{TR}$, in that order, facilitating a strong bi-frequency PM response and thus flow response (at $f_\mathrm{TR},f_\mathrm{TR2}$) compared to an essentially multi-frequency response (at $f_\mathrm{TR},f_\mathrm{PB}$) for other PMs ($m_\mathrm{UC1-3}$). This outcome suggests a few features of the PM-FSI system. First, natural frequencies nearer to the inherent flow timescale of $1/f_{\text{VS}}$ are more likely to be amplified than those further away from it (e.g., at $f = f_{\text{VS}}/2$). Second, the frequency protection of the truncation resonance, along with its capacity to isolate the structural response to the flow-surface interface mass, promotes a stronger interplay than a pass band mode. For example: at $\lambda_1$, the PM with $m_{\text{UC}1}$ has a pass band frequency $f_{\text{PB}}$ near to the truncation resonance frequency $f_{\text{TR}2}$ associated with $m_{\mathrm{UC}5}$. However, it is the latter PM that provides the more dominant response at that frequency. This behavior suggests that designing truncation resonances to align with flow behavior could be a key imperative to effectively modulating flows with PMs.

For a higher displacement envelope, $\lambda_2$, the mechanical energy in the PM and the flow concentrates significantly in the primary eigenmode, $f_\mathrm{TR}$, for all unit cell masses ($m_\mathrm{UC1-5}$), although a weak broadband nature of the PM behavior persists (Fig.~\ref{fig:FFT_CBars}a.i). In addition to the primary harmonic, significant peaks at the second harmonic, $2f_\mathrm{TR}=f_\mathrm{VS}$, are also visible in the $F_\mathrm{CS}$, and $C_{l}$ signals (Fig.~\ref{fig:FFT_CBars}a.(ii-iii)). 

Further increasing the displacement envelope to $\lambda_3$ ($\rightarrow\lambda_\mathrm{high}$) results in a clear, narrow-band peak at the prescribed $f_\mathrm{TR}$ in both the PM and the flow, with a strong second-harmonic generation in the flow. Moreover, a similar frequency spectrum for both the PM and flow response for all values of $m_\mathrm{UC}$ indicates that the unit cell mass plays a negligible role in determining the FSI dynamics. That is, for higher displacement amplitudes $\lambda$, the FSI dynamics more strongly synchronize onto the aligned timescale, and the broader band behavior associated with tuning $m_{\mathrm{UC}}$ is less impactful to the coupled dynamics.

\subsubsection{FSI Dynamics: $f_\mathrm{TR}=f_\mathrm{VS}$}

Now, consider the PMs featuring a truncation resonance aligned with the global attractor frequency of the rigid body flow, i.e., $f_\mathrm{TR}=f_\mathrm{VS}$ (Fig.~\ref{fig:FFT_CBars}b). Though the PM and flow behavior are broadband at $\lambda_1$, the spectral peak at the prescribed $f_\mathrm{TR}$ is relatively stronger than those at the pass band frequencies, $f_\mathrm{PB}$. Note that a second truncation resonance, $f_\mathrm{TR2}$, does exist for these PMs; however, they play no role in the FSI dynamics. Similar to the results for $f_\mathrm{TR}=0.5f_\mathrm{VS}$, as the displacement envelope increases ($\lambda_1\rightarrow\lambda_3$), the PM exhibits a strong narrow-band response, centered at a frequency nearly equal to $f_\mathrm{TR}$ for all values of $ m_\mathrm{UC}$. The PM dynamics are subsequently replicated in the interface variable, $F_\mathrm{CS}$, along with second harmonic generation of the excited frequencies. However, none of these second harmonics survive for long in the flow, as seen in the $C_{l}$ signal, which measures the lift force along the entire inclined flat plate surface, as they lie far away from $f_\mathrm{VS}$. In addition, from Fig.~\ref{fig:FFT_CBars}b, results show the fluid-PM coupling is relatively strong and narrow-band $f\approx f_\mathrm{TR}$ in the $f_\mathrm{TR}=f_\mathrm{VS}$ case, compared to a bi-frequency coupling at $f\approx f_\mathrm{TR},2f_\mathrm{TR}$ in the $f_\mathrm{TR}=0.5f_\mathrm{VS}$ case. This stronger synchrony is due to the alignment of the primary harmonic with the global rigid body flow attractor, and the smaller range of displacement envelope, $\lambda_\mathrm{Range}$, available for $f_{\mathrm{TR}}= f_{\mathrm{VS}}$ compared with $f_{\mathrm{TR}} = 0.5f_{\mathrm{VS}}$ (see Tab.~\ref{tab:FSI_Sims}).

\subsubsection{FSI Dynamics: $f_\mathrm{TR}=2f_\mathrm{VS}$}

Finally, Fig.~\ref{fig:FFT_CBars}c plots the spectral frequency composition of the PM dynamics for PMs with $f_\mathrm{TR}=2f_\mathrm{VS}$, normalized by the maximum spectral amplitude observed in the 15 FSI configurations investigated for this $f_\mathrm{TR}$. The PMs exhibit a broadband response in all FSI configurations with significant excitation of both the prescribed truncation resonance and the pass band frequencies. However, though not directly apparent from Fig.~\ref{fig:FFT_CBars}c, the PMs in all 15 FSI configurations exhibit dynamics an order to two orders of magnitude ($O(10^{1-2})$) smaller when compared to PM dynamics with $f_\mathrm{TR}=0.5f_\mathrm{VS}$, and $f_\mathrm{TR}=f_\mathrm{VS}$. 

Fig.~\ref{fig:Rigid_vs_PM}b compares the spectral compositions of the PM dynamics in one of these 15 FSI configurations of PMs with $f_\mathrm{TR}=2f_\mathrm{VS}$, with spectral amplitudes normalized by the maximum amplitude seen in a representative case chosen from the 15 FSI simulations of PMs with $f_\mathrm{TR}=f_\mathrm{VS}$. The relatively small spectral peaks of the PM with $f_\mathrm{TR}=2f_\mathrm{VS}$ corroborate the order of magnitude difference in dynamics between the different PMs. Specifically, the muted response in the former PM, stems from the `excitable' flow frequencies ($f\leq f_\mathrm{VS}$) lying entirely in the primary band gap for PMs with $f_\mathrm{TR}=2f_\mathrm{VS}$, thus, having a negligible effect on the flow, and rendering the flow behavior virtually identical to that of the rigid inclined flat plate configuration. In addition, note that all FSI configurations investigated in this article exhibit a non-zero mean displacement, $\chi_\mathrm{mean}>0$, at the interface, different from the zero displacement of the rigid inclined flat plate. However, the virtually unchanged flow dynamics for $f_\mathrm{TR}=2f_\mathrm{VS}$, compared with the rigid case, indicate that the PM dynamics ($\chi_\mathrm{amp}>0$) are critical to alter the flow.

\subsubsection{FSI Dynamics: Nonlinear Frequency Shift at high $\lambda$}\label{subsubsec:NL_FSI}

We encounter another interesting phenomenon in the strongly coupled FSI configurations ($\lambda_3$) in Fig.~\ref{fig:FFT_CBars}b.i. The narrow band PM response concentrates at a frequency that is slightly lower in magnitude than the prescribed $f_\mathrm{TR}$. This is more clearly visible in the inset in Fig.~\ref{fig:Rigid_vs_PM}b.i, where the peak frequency shifts to the left of $f_\mathrm{TR}(=f_\mathrm{VS})$. Although a detailed analysis lies outside the scope of this article, we hypothesize that a non-linear phenomenon arising from the strong coupling between the aerodynamic flow and the PM at the fluid-PM interface is responsible for the frequency shift. As the PM exhibits high surface displacement and couples strongly with the oncoming flow, a fluid-added mass essentially latches onto the interface PM mass, $m_1$, altering the modal characteristics of the PM. 

\subsection{Summary of Static and Dynamic PM-FSI behaviors}\label{subsec:FSI_S&D}

\begin{figure*}[t!]
    \centering
    \includegraphics[width=\textwidth]{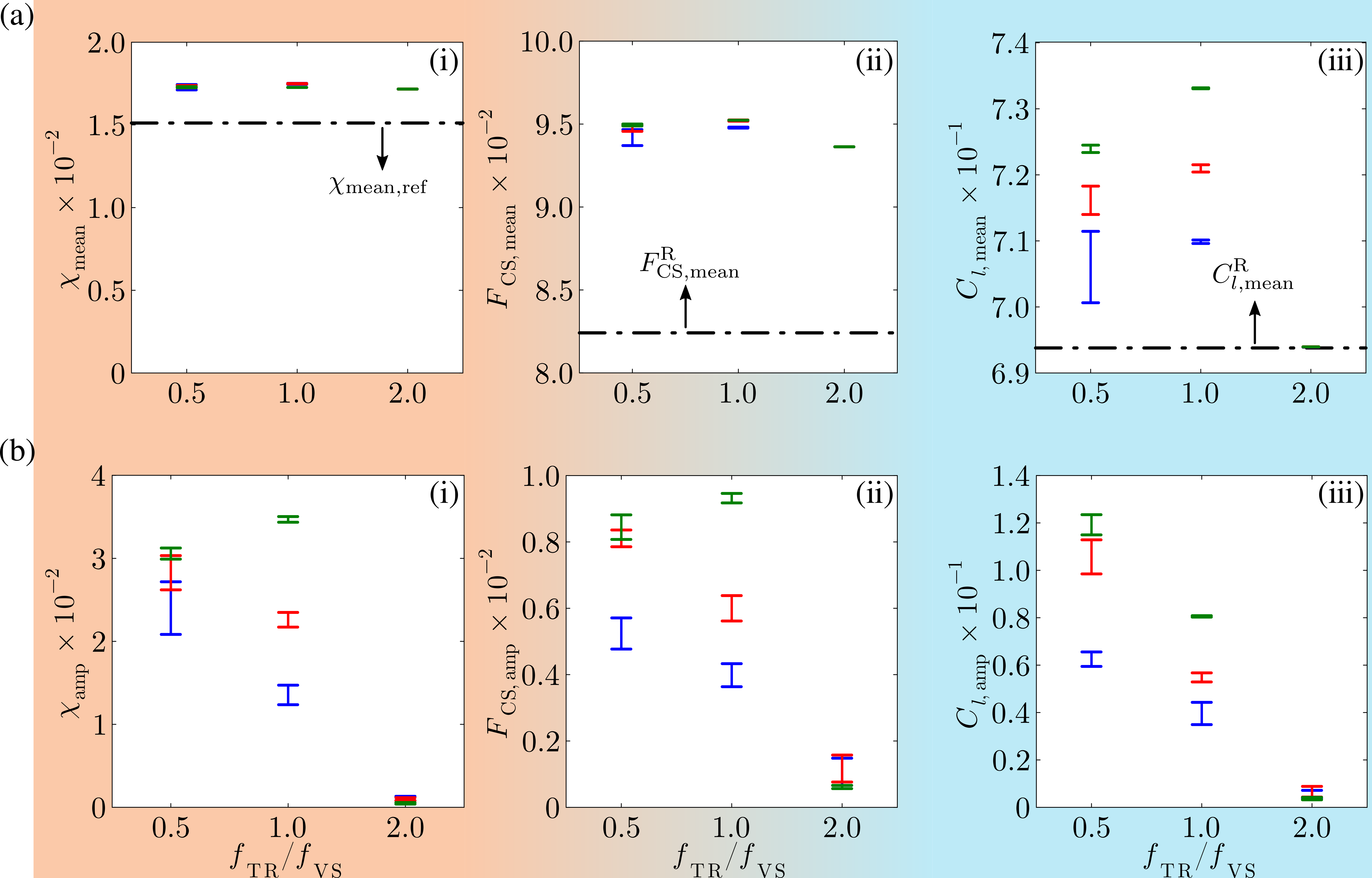}
    \caption{Steady and Dynamic PM-FSI parameters. (a) Time-average (mean), and (b) dynamic amplitude of (i) $\chi$, (ii) $F_{\mathrm{CS}}$, (iii) and $C_{l}$. Colors indicate different displacement amplitudes $\lambda$, increasing from blue (small) through red (intermediate) to green (large), and the range indicated by the bar corresponds to results across $m_{\mathrm{UC}, \text{Range}}$ for that $\lambda$, $f_{\mathrm{TR}}$ value. Reference values in the dot-dash lines are for (i) the target mean value chosen for (i) $\chi_\mathrm{mean}$, and (ii), (iii) the values - $F_\mathrm{CS,mean}^\mathrm{R}$, and $C_{l,\mathrm{mean}}^\mathrm{R}$, for the rigid body configuration (i.e., $\Gamma_c = \{0\}$).}
    \label{fig:Qualitative_data}
\end{figure*}

Our aim in this subsection is to clarify the role of static versus dynamic structural behavior in modulating the flow, to provide a distinct but compatible perspective to the analysis of the dynamic frequency content shown in the prior subsections, and to connect the various PM-FSI dynamics to changes in aerodynamic quantities of interest (mean lift).

Fig.~\ref{fig:Qualitative_data} summarizes mean and dynamic amplitude content for the 45 different PM parameters considered. Fig.~\ref{fig:Qualitative_data}(a.i) shows that the mean displacement $\chi_\mathrm{mean}$ is nearly equivalent for all PM cases considered. (Consistent with this observation, $F_{\mathrm{CS}, \text{mean}}$ is nearly identical across all PM parameter choices). This result confirms that the effective stiffness $k_\mathrm{eff}$ is responsible for establishing this static displacement. This value of displacement is slightly different from the target value determined using $F_{\mathrm{CS}}^\mathrm{R}$ associated with the rigid configuration, reflective of the fact that under FSI coupling the mean load on the compliant surface is different from the rigid case. However, the mean displacement remains within approximately $1.5\%$ of the target value. More than that, the static displacement of the structure does not contribute to either the dynamic content in the FSI system, or to changes in mean lift. To see this fact, note that for $f_\mathrm{TR} = 2f_\mathrm{VS}$, there is essentially no dynamic content (the dynamic amplitudes in $\chi$ and $F_\mathrm{CS}$ are negligible), as seen in Fig. \ref{fig:Qualitative_data}b. In this case, despite the nonzero mean displacement, the mean lift remains nearly unchanged from the rigid reference case. Because the other PM cases, associated with other values of $f_{\mathrm{TR}}$, have the same mean value of displacement and $F_\mathrm{CS}$, the changes to their dynamic content and to mean lift must be induced directly by the dynamics. This demonstrates the importance of $k_{\mathrm{eff}}$ in setting the static displacement of the structure, and of the other parameters in driving the dynamics and changes to aerodynamic lift.

FSI dynamics become more pronounced for $f_\mathrm{TR}\ne 2f_\mathrm{VS}$, with corresponding increases to $C_{l,\mathrm{mean}}$, relative to the rigid case, of $1.3-4.6\%$ for $f_\mathrm{TR}=0.5f_\mathrm{VS}$ and $2.5-5.7\%$ for $f_\mathrm{TR}=f_\mathrm{VS}$. At a given value of $f_\mathrm{TR}$, the largest mean lift occurs for the largest displacement amplitude $\lambda$ as observed in Fig. \ref{fig:Qualitative_data}(a.iii). This outcome indicates that greater FSI synchrony, with sharper frequency content, is conducive to larger mean lift benefits. Moreover, at these larger $\lambda$ values, the effect of $m_\mathrm{UC}$ on mean lift is negligible. This is consistent with the frequency analysis from before, where $m_\mathrm{UC}$ is only a relevant parameter for low $\lambda$ values where the dynamics are broader band. The spread in the range of mean lift values observable for low $\lambda$ values demonstrates that the different broadband dynamics obtainable by tuning $m_\mathrm{UC}$ produce differences in the mean lift. 

\subsection{FSI results justify the proposed PM behavioral parameters}

Overall, simulation results show a wide range of FSI dynamics, with distinct flow vortex-shedding characteristics, e.g., frequency, spatial wavelength, and intensity (vorticity) when systematically varying the truncation resonance frequency, $f_\mathrm{TR}$, the displacement envelope, $\lambda$, and the unit cell mass, $m_\mathrm{UC}$ of the diatomic PM. Additionally, even though the effective stiffness, $k_\mathrm{eff}$, is constant across all FSI simulations, we can still draw important inferences from the FSI results in Sec.~\ref{subsec:FSI_fTR}-\ref{subsec:FSI_S&D}, and justify the choice of $k_\mathrm{eff}$ as a critical PM behavioral parameter that governs the FSI dynamics. 

The behavioral parameter $k_\mathrm{eff}$ is critical to ensuring comparable FSI results with different PM dynamics, and sets the bounds of other behavioral parameters. In our simulations, we chose $k_\mathrm{eff}$ based on prescribed surface motion simulations, such that the static displacement of the compliant section had a negligible impact on the surrounding aerodynamic flow. Note that reducing $k_\mathrm{eff}$ to achieve higher mean static surface displacements outside the targeted range would initiate new types of nonlinear fluid-PM interactions, making the FSI behavior more complex. In addition, $k_\mathrm{eff}$ bounds the PM dynamics. For a given $k_\mathrm{eff}$, the upper limit ($\lambda_\mathrm{high}$) of the customizable PM displacement envelope range, $\lambda_\mathrm{Range}$, is determined by the $\lambda_\mathrm{s}$ of a single mass-spring oscillator with a spring stiffness, $k_\mathrm{s}=k_\mathrm{eff}$. Therefore, PM models with a suitable $k_\mathrm{eff}$ are vital to engineering the ensuing FSI.

The truncation resonance frequency, $f_\mathrm{TR}$, tunes the targeted dominant (first mode) structural behavior relative to the vortex-shedding frequency. This is perhaps the most obvious behavioral parameter choice. This temporally distinct dynamic response translates to a wide variety of flow vortex-shedding characteristics. The FSI results show distinct dynamics of the coupled system as a function of truncation resonance frequency, resulting in either strong frequency signatures around the truncation resonance, at its second harmonic, or no dynamics. This is also partially due to the combined parameters of the PM, which fully dictate whether the PM response is single-, double-, or multi-frequency response to the input broadband fluid force.  

The resonant displacement envelope, $\lambda$, dictates the PM response amplitude that, in turn, dictates the fluid-PM coupling strength at the interface. This coupling governs the strength of the vortex-shedding process, and whether second harmonic (and, more broadly, multi-frequency) generation appears in the FSI system.  For example, the FSI results for the $f_\mathrm{TR}=0.5f_\mathrm{VS}$ cases show that a small displacement envelope leads to a broadband PM response with active participation of other structural resonances such as the pass band modes and the $2^\mathrm{nd}$ truncation resonance in the PM dynamics. In some cases, the modal participation of these modes is higher than the engineered primary truncation resonance mode, leading to a broadband and low-intensity vortex-shedding behavior in the flow. Alternatively, for a high displacement envelope, the PM dynamics are localized at the interface and concentrated at a frequency very near to the primary truncation resonance frequency. The spectral contents of the flow in these scenarios are commensurate with the PM dynamics, with additional participation from the nonlinearly-generated $2^\mathrm{nd}$ harmonics.

Finally, the unit cell mass, $m_\mathrm{UC}$, determines the frequency components excited in low $\lambda$ PMs, where the FSI dynamics are broader band. For example, a smaller unit cell mass causes different frequencies to be activated in the PM, interface, and flow. By contrast, the unit cell mass plays a negligible role in the FSI dynamics for high $\lambda$ PMs, where dynamics are localized near a small number of frequencies driven by the relationship between $f_\mathrm{TR}$ and $f_{\mathrm{VS}}$. In addition, $m_\mathrm{UC}$ partially determines the total absolute mass of the diatomic PM, which will influence the operational considerations of the aircraft, e.g., fuel costs, wing design to integrate the PM, etc. Therefore, $m_\mathrm{UC}$ also has practical justifications in the context of a given aerodynamic flow. 

\section{Conclusions}\label{sec:Conclusion}

This study systematically explores the fundamental FSI mechanisms responsible for changes in flow characteristics in a canonical PM-FSI setting, complementing existing studies and motivating the exploration of future FSI configurations with different PMs. We identify that grounded PM models are critical in aerodynamic FSI configurations, since lift preferentially activates the first mode of the structure, and grounded PMs enable a truncation resonance at the first structural mode. We show that representative flow forces can activate the truncation resonance in grounded PMs but not in ungrounded PMs, where the truncation resonance is a higher-order mode. We introduce the concept of PM behavioral parameters that relate specific PM behaviors, centered around a targeted truncation resonance, with specific FSI-relevant flow effects. The idea of viewing PM models through the lens of critical behavioral parameters as opposed to the traditional approach of viewing PMs based on structural parameters, or structural-driven normalizations, presents a new perspective on selecting effective PM models for a given flow configuration.

The four critical PM behavioral parameters we identify are effective stiffness, truncation resonance frequency, unit cell mass, and the resonant displacement envelope. These parameters can be tuned to dictate the steady and dynamic (spectral) characteristics of the PM-FSI system. We also provide an unambiguous mapping that relates targeted behavioral parameters to a unique set of structural parameters, from which a simulation or experiment can be performed.  We perform high-fidelity, fully coupled numerical FSI studies of various PM models with different target behavioral parameters, revealing distinct flow behavior. The effective stiffness determines the mean static displacement at the fluid-PM interface, which is found to have a negligible impact on the flow behavior. However, the effective stiffness also dictates the customization ranges of the displacement envelope and the unit cell mass, which has an indirect influence on the FSI. The truncation resonance frequency determines the dominant frequencies excited in the PM-FSI setting when engineered in the proximity of the inherent flow instability frequency, $f_\mathrm{VS}$. For $f_\mathrm{TR}<f_\mathrm{VS}$, depending on other PM behaviors, we observe a single, double-, or multi-frequency FSI dynamics. Alternatively, for $f_\mathrm{TR}>f_\mathrm{VS}$, we observe negligible FSI dynamics, leaving the flow behavior virtually unaltered as compared to the rigid baseline case. Both the displacement envelope and unit cell mass dictate how broadband the FSI dynamics are, and therefore, these parameters emerge as important FSI parameters from this study. The displacement envelope emerges as a powerful PM control parameter that dictates how synchronized the dynamics are around the tuned relationship between the truncation resonance frequency and the inherent flow frequency. These effects drive the intensity of the vortex shedding process and lead to nonlinear frequency modulation. It further highlights a new finding of this article: absolute amplitudes of PMs have thus far not been considered in the PM literature, but our results show this is a critical parameter in the context of FSI. Finally, for parameters where the displacement amplitude is small, resulting in lower amplitude coupling (i.e., smaller dynamic amplitudes of the structural motion and lift), the unit cell mass defines the locations of the other natural frequencies, which conspire with the flow to produce the specific broadband content for that parameter set.

We observe a maximum net increase in the mean lift force, $\Delta C_{l,\mathrm{mean}} \%\approx 5.7\%$, in FSI settings featuring PMs with high $\lambda$ values, when compared to the rigid inclined flat plate. While not the focus of this study, this result indicates potential benefits of PM-based FSI strategies in improving the operational efficiency of aerial vehicles. The pursuit of this aim is facilitated through the proposed PM behavioral parameters, which were constructed to be generalizable to diverse flow settings with prominent frequency content, and to more complex PMs such as auxetic lattices~\cite{ZhangAFM2025}, topological PMs~\cite{VishwakarmaCP2025}, and nonlinear lattices~\cite{PatilAM2022}. Applying the proposed behavioral parameter framework to these configurations presents a promising avenue for exploring novel PM-FSI configurations.

\section*{CRediT authorship contribution statement}
Vinod Ramakrishnan (VR): investigation, formal analysis, data curation, visualization, writing – original draft, review, and editing. Arturo Burgos (AB): investigation, formal analysis, data curation, writing – original draft, and editing. Sangwon Park (SP): investigation, formal analysis. Kathryn Matlack (KM): conceptualization, supervision, formal analysis, funding acquisition, writing–review, and editing. Andres Goza (AG): conceptualization, supervision, formal analysis, funding acquisition, writing–review, and editing. 

\section*{Data availability} 
Data will be made available on request. 

\section*{Acknowledgments} 
This material is based upon work supported by the Air Force Office of Scientific Research under \red{award number FA9550-21-1-0182 and award number FA9550-23-1-0299}, and the Grainger College of Engineering at the University of Illinois Urbana-Champaign through the \red{Strategic Research Initiative}.

\section*{Appendix}

\setcounter{equation}{0}
\renewcommand{\theequation}{A\arabic{equation}}
\renewcommand{\theHequation}{A\arabic{equation}}
\setcounter{figure}{0}
\renewcommand{\thefigure}{A\arabic{figure}}
\renewcommand{\theHfigure}{A\arabic{figure}}
\setcounter{section}{0}
\renewcommand{\thesection}{A\arabic{section}}
\renewcommand{\theHsection}{A\arabic{section}}
\setcounter{table}{0}
\renewcommand{\thetable}{A\arabic{table}}
\renewcommand{\theHtable}{A\arabic{table}}

\section{Mass and Stiffness Matrices of the Diatomic PM }\label{sec:MK_Matrices}

The non-dimensional mass and stiffness matrices collating the equations of motion of all $N$ unit cells of a general diatomic PM are given as:

\begin{equation*}
    \mathbf{M}=\frac{\hat{\mathbf{M}}}{\rho_\mathrm{fluid}hl}=\begin{bmatrix}
                    m_1 & 0 & 0 & \cdots & 0\\
                    0 & m_2 & 0 & \cdots & 0\\
                    0 & 0 & m_1 & \ddots & \vdots\\
                    \vdots & \vdots & \ddots & \ddots & 0\\
                    0 & 0 & \cdots & 0 & m_2
                \end{bmatrix}_{2N\times2N},
\end{equation*}
\begin{equation*}
    \mathbf{K}=\frac{\hat{\mathbf{K}}}{\rho_\mathrm{fluid}U^2}=\begin{bmatrix}
                    k_1 & -k_1 & 0  & \dots & 0\\
                    -k_1 & k_1+k_2+k_\mathrm{g} & -k_2 & \dots & 0\\
                    0 & -k_2  & k_1+k_2 & \ddots & \vdots\\
                    \vdots & \vdots & \vdots & \ddots  & -k_2\\
                    0 & 0 & \dots & -k_2 & k_2+k_\mathrm{g}
                \end{bmatrix}_{2N\times2N}.
\label{eq:MK_Matrices}
\end{equation*}

\section{Additional Interface Operators}\label{sec:OperatorY_&_Y_tilde}

We define the load-assembly operator $\widetilde{\mdisc{Y}}$, which combines
normal projection, restriction to $\Gamma_c$, quadrature summation, and
injection into the first PM load entry, providing the discrete counterpart of Eqns.~\eqref{eq:f_pm}--\eqref{eq:F_Gamma_c}. Examining the right–hand side of Eq.~\eqref{eqn:mass_spring_sd}, applying $\widetilde{\mdisc{Y}}$ to the spatially discrete surface–stress field  $(\mathbf{f}=(\mathbf{f}_1,\ldots,\mathbf{f}_{N_\Gamma})$ with $\mathbf{f}_j\in\mathbb{R}^2$ gives
\begin{equation}
\big(\widetilde{\mdisc{Y}}\,\mathbf{f}\big)_k
\;=\;
\delta_{k1}\sum_{j=1}^{N_\Gamma} d_j\,\hat{\mathbf n}_j^{\!\top}\mathbf{f}_j,
\qquad
k=1,\ldots,2N,
\label{eq:Ytilde_action_indicator}
\end{equation}
where
\begin{equation}
d_j \;\coloneqq\; \sum_{m\in I_c}\delta_{jm}
\;=\;
\begin{cases}
1, & j\in I_c,\\
0, & \text{otherwise},
\end{cases}
\qquad j=1,\ldots,N_\Gamma
\label{eq:Sc_indicator}
\end{equation} and ${I_c}$ is the index set of body points on the compliant section $\Gamma_c$.

We define the kinematic operator $\mdisc{Y}$ by first masking the compliant
body points via $d_j$, then applying the Gaussian envelope $g_i(\chi)$, and finally scale by the interface rate $\dot{\chi} = \xi = \bf e_1^\top \bm{\xi}$, orienting the result along the local unit normal $\hat{\mdisc{n}}_j$ at each body point in $\Gamma_c$:

\begin{equation}
\big(\mdisc{Y}\,\bm{\sdisc{\xi}}\big)_j
\;\coloneqq\;
d_j\,g_j(\chi)\,
\big(\mathbf{e}_1^{\!\top}\bm{\sdisc{\xi}}\big)\,
\hat{\mathbf n}_j,
\qquad j=1,\ldots,N_\Gamma.
\label{eq:Y-action}
\end{equation} Here
\begin{equation}
g_j(\chi)\;\coloneqq\;\exp\!\left(-\frac{\big[3\big(2\,s(\tilde s_j,t)-1\big)\big]^2}{2}\right),
\label{eq:mask-and-gaussian}
\end{equation} with $s(\tilde s_j,t)$ the inverse of the stretch map $\tilde s=\beta\,f(s;\alpha)$,
and $\hat{\mathbf n}_j$ the unit normal at body point $j$.

\section{Grid Convergence}\label{sec:grid_convergence}

We verify that the numerical parameters used in the main text are adequate by the comparisons in Table~\ref{tab:grid}. The grid-convergence check varies (i) Eulerian grid spacing, (ii) total domain size (via the number of multi-domain levels). For each case, the time step $\Delta t$ is chosen so that the Courant–Friedrichs–Levy (CFL) number is $\approx 0.2$. The chosen Lagrangian body-point spacing to the Eulerian spacing is $\Delta s/ \Delta x = 2$.  Finally the smallest sub-domain for all cases is $[-1,2]\times[-1,1]$. Grid~2 is the setup used for the production simulations. The last columns of the table report the percentage variations in the mean lift coefficient, $C_{l,\mathrm{mean}}$, and in the vortex-shedding frequency relative to Grid~1 (most refined grid), thereby providing a measure of the sensitivity of both mean and unsteady quantities to the numerical parameters. 

\begin{table}[!htp]
\centering
\setlength{\tabcolsep}{4pt}
\begin{tabular}{lccccccc}
\hline
Grid & $\Delta x$ & $\Delta t$ & Total domain & $C_{l, \mathrm{mean}}$ & \%$|\delta C_{l,\mathrm{mean}}|$ & $f_\mathrm{VS}$  &  \%$|\delta f_\mathrm{VS}|$\\
\hline
7 & 0.0400 & 0.0080  & $[-5.5,6.5]\times[-4,4]$   & 0.666 & 5.128 & 0.5202 & 18.104\\ 
6 & 0.0200 & 0.0040  & $[-5.5,6.5]\times[-4,4]$   & 0.691 & 1.567 & 0.5953 & 6.281\\
5 & 0.0100 & 0.0020 & $[-5.5,6.5]\times[-4,4]$  & 0.675 & 3.846 & 0.6090 & 4.125\\
4 & 0.0050 & 0.0010  & $[-23.5,24.5]\times[-16,16]$   & 0.676 & 3.704 & 0.6211 & 2.2197\\
3 & 0.0050 & 0.0010  & $[-11.5,12.5]\times[-8,8]$   & 0.681 & 2.991 & 0.6181 & 2.692\\
2* & 0.0050 & 0.0010  & $[-5.5,6.5]\times[-4,4]$ & 0.694 & 1.139  & 0.6256 & 1.511\\
1 & 0.0025 & 0.0005  & $[-5.5,6.5]\times[-4,4]$  & 0.702 & N/A & 0.6352 & N/A\\
\hline
\end{tabular}
\caption{Grid/domain/time-step refinement at $\theta=12^\circ$. Grid~2* is the production setup; percentage change is relative to Grid~1.}
\label{tab:grid}
\end{table}

\section{PM Structural Parameters for FSI Simulations}

Tab.\ref{tab:FSI_Sims} provides the 45 distinct PM structural and behavioral parameters explored for FSI simulations in this article. Note that all PMs have an identical effective stiffness, $k_\mathrm{eff}=5.4533$.

\renewcommand{\arraystretch}{1.5}
\begin{table}[t!]
    \centering
    \begin{tabular}{|M{1cm}|M{1cm}|M{2cm}|M{2cm}|M{2cm}|M{2cm}|M{2cm}|}\hline
    \multicolumn{7}{|c|}{$\bm{f_\mathrm{TR}=0.5f_\mathrm{VS}=0.3128}$}\\ \hline
     \multicolumn{2}{|c|}{\multirow{2}*{$\bm{\{m_1,m_2,k,k_\mathrm{g}\}}$}}& $\bm{m_\mathrm{UC1}}$ & $\bm{m_\mathrm{UC2}}$ & $\bm{m_\mathrm{UC3}}$ &$\bm{m_\mathrm{UC4}}$ & $\bm{m_\mathrm{UC5}}$\\ \cline{3-7}
     \multicolumn{2}{|c|}{} & $\bm{1.39}$ & $\bm{1.876}$ & $\bm{2.953}$ & $\bm{4.759}$ & $\bm{7.573}$\\ \hline
    $\bm{\lambda_1}$ & $\bm{0.054}$ & $\{0.224,1.164$, $13.23,5.424\}$ & $\{0.346,1.529$, $10.77,7.431\}$ & $\{0.472,2.48$, $8.5,11.9\}$ & $\{0.583,4.176$, $7.19,19.413\}$ & $\{0.695,6.878$, $6.49,31.152\}$\\ \hline
    $\bm{\lambda_2}$ & $\bm{0.107}$ & $\{0.777,0.613$, $11.83,6.388\}$ & $\{0.819,1.057$, $9.83,8.749\}$ & $\{0.866,2.086$, $8.02,13.794\}$ & $\{0.931,3.827$, $6.95,22.24\}$ & $\{1,6.572$, $6.36,35.234\}$\\ \hline
    $\bm{\lambda_3}$ & $\bm{0.161}$ & $\{1.252,0.138$, $8.59,11.597\}$ & $\{1.234,0.642$, $7.87,14.56\}$ & $\{1.225,1.728$, $7.04,21.12\}$ & $\{1.236,3.523$, $6.46,31.912\}$ & $\{1.262,6.312$, $6.1,48.678\}$\\ \hline 
    \end{tabular}\\[0.2cm]
    \begin{tabular}{|M{1cm}|M{1cm}|M{2cm}|M{2cm}|M{2cm}|M{2cm}|M{2cm}|}\hline
    \multicolumn{7}{|c|}{$\bm{f_\mathrm{TR}=f_\mathrm{VS}=0.6256}$}\\ \hline
     \multicolumn{2}{|c|}{\multirow{2}*{$\bm{\{m_1,m_2,k,k_\mathrm{g}\}}$}}& $\bm{m_\mathrm{UC1}}$ & $\bm{m_\mathrm{UC2}}$ & $\bm{m_\mathrm{UC3}}$ &$\bm{m_\mathrm{UC4}}$ & $\bm{m_\mathrm{UC5}}$\\ \cline{3-7}
     \multicolumn{2}{|c|}{} & $\bm{0.347}$ & $\bm{0.469}$ & $\bm{0.738}$ & $\bm{1.190}$ & $\bm{1.893}$\\ \hline
    $\bm{\lambda_1}$ & $\bm{0.107}$ & $\{0.059,0.288$, $13.23,5.424\}$ & $\{0.088,0.38$, $10.77,7.431\}$ & $\{0.118,0.619$, $8.5,11.9\}$ & $\{0.145,1.045$, $7.19,19.413\}$ & $\{0.172,1.72$, $6.49,31.087\}$\\ \hline
    $\bm{\lambda_2}$ & $\bm{0.215}$ & $\{0.195,0.152$, $11.83,6.388\}$ & $\{0.204,0.265$, $9.83,8.749\}$ & $\{0.217,0.52$, $8.02,13.794\}$ & $\{0.233,0.957$, $6.95,22.24\}$ & $\{0.251,1.643$, $6.36,35.298\}$\\ \hline
    $\bm{\lambda_3}$ & $\bm{0.322}$ & $\{0.313,0.034$, $8.59,11.597\}$ & $\{0.308,0.16$, $7.88,14.499\}$ & $\{0.306,0.432$, $7.04,21.12\}$ & $\{0.31,0.88$, $6.46,32.106\}$ & $\{0.315,1.577$, $6.1,48.617\}$\\ \hline 
    \end{tabular}\\[0.2cm]
    \begin{tabular}{|M{1cm}|M{1cm}|M{2cm}|M{2cm}|M{2cm}|M{2cm}|M{2cm}|}\hline
    \multicolumn{7}{|c|}{$\bm{f_\mathrm{TR}=2f_\mathrm{VS}=1.2512}$}\\ \hline
    \multicolumn{2}{|c|}{\multirow{2}*{$\bm{\{m_1,m_2,k,k_\mathrm{g}\}}$}}& $\bm{m_\mathrm{UC1}}$ & $\bm{m_\mathrm{UC2}}$ & $\bm{m_\mathrm{UC3}}$ &$\bm{m_\mathrm{UC4}}$ & $\bm{m_\mathrm{UC5}}$\\ \cline{3-7}
    \multicolumn{2}{|c|}{} & $\bm{0.087}$ & $\bm{0.117}$ & $\bm{0.185}$ & $\bm{0.297}$ & $\bm{0.473}$\\ \hline
    $\bm{\lambda_1}$ & $\bm{0.215}$ & $\{0.015,0.072$, $13.23,5.424\}$ & $\{0.022,0.095$, $10.77,7.431\}$ & $\{0.03,0.154$, $8.52,11.843\}$ & $\{0.036,0.261$, $7.2,19.368\}$ & $\{0.043,0.43$, $6.49,31.087\}$\\ \hline
    $\bm{\lambda_2}$ & $\bm{0.429}$ & $\{0.049,0.038$, $11.74,6.457\}$ & $\{0.051,0.065$, $9.87,8.686\}$ & $\{0.054,0.13$, $8.02,13.794\}$ & $\{0.058,0.238$, $6.96,22.202\}$ & $\{0.063,0.41$, $6.36,35.234\}$\\ \hline
    $\bm{\lambda_3}$ & $\bm{0.644}$ & $\{0.078,0.008$, $8.63,11.478\}$ & $\{0.077,0.04$, $7.88,14.499\}$ & $\{0.077,0.108$, $7.04,21.12\}$ & $\{0.077,0.219$, $6.46,32.042\}$ & $\{0.079,0.394$, $6.1,48.617\}$\\ \hline 
    \end{tabular}
    \caption{FSI PM behavioral and structural parameters. ($k_\mathrm{eff}=5.4533$)}
    \label{tab:FSI_Sims}
\end{table}

The PM structural parameters in Tab.~\ref{tab:FSI_Sims}, were obtained by the optimization procedure described in Sec.~\ref{subsec:PM_Behaviors}. We can draw a few observations from the above parameter sets:
\begin{itemize}
    \item The unit cell masses and the available customization range, $m_\mathrm{UC1-5}\in m_\mathrm{UC,Range}$, scale with $1/f_\mathrm{TR}^2$, consistent with $f\propto1/\sqrt{m}$.
    \item The displacement envelopes and the available customization range, i.e., $\lambda_{1-3}\in\lambda_\mathrm{Range}$, scale with $f_\mathrm{TR}$. This is also an intuitive effect as $\lambda$ is associated with units of velocity. So, given a displacement amplitude, $\chi_\mathrm{amp}$, the velocity amplitude, $\dot{\chi}\propto2\pi f\chi_\mathrm{amp}$, as the velocity is a time derivative of displacement ($\dot{\chi}=\frac{d\chi}{dt}$).
    \item The mass ratio, $m_\mathrm{r}=m_2/m_1$, remains constant for a proportional $\{\lambda,m_\mathrm{UC}\}$ in the respective ranges, across different $f_\mathrm{TR}$'s, with only the absolute mass changing to produce the desirable PM behaviors for different prescribed $f_\mathrm{TR}$'s. For example, $\{\lambda_2,m_\mathrm{UC3}\}: \ m_\mathrm{r}=\frac{2.086}{0.866}=\frac{0.52}{0.217}=\frac{0.13}{0.054}\approx2.4$.
    \item The absolute stiffnesses, $k_\mathrm{g}$, and $k$, and consequently, the grounding stiffness ratio, $k_\mathrm{gr}$, both remain constant for a given $\{\lambda,m_\mathrm{UC}\}$ across different $f_\mathrm{TR}$'s. For example, $\{\lambda_2,m_\mathrm{UC3}\}: \ \{k,k_\mathrm{g}\}=\{8.02,13.794\}$.
\end{itemize}

\end{document}